\documentclass{jkas}
\usepackage{kotex}
\usepackage{color}
\usepackage{soul}
\def\year{2018} % publication year
\def\month{April} % publication month
\def\volume{00} % journal volume
\def\issue{0} % journal issue
\def\beginpage{1} % first page of article
\def\endpage{99} % last page of article
\def\received{---} % date paper was received by JKAS
\def\accepted{---} % date of acceptance
\date{Received \received; accepted \accepted}

\title{
Correlation study of temporal and emission properties of quiescent magnetars
}

\author[1]{Jiwoo Seo}
\author[1,2]{Jaewon Lee}
\author[1]{Hongjun An}
\affil[1]{Department of Astronomy and Space Science, Chungbuk National University, Cheongju 28644, Republic of Korea \email{hjan@cbnu.ac.kr}}
\affil[2]{METASPACE, Cheongju 28156, Republic of Korea \email{jwlee@metaspace.co.kr}}

\def\corrauthor{
Jaewon Lee; jwlee@metaspace.co.kr
Hongjun An; hjan@cbnu.ac.kr
}

\def\runningauthor{Seo et al.
}

\def\runningtitle{
Correlation study of magnetars' emission properties
}

\def\keywords{
stars: neutron --- stars: statistics --- pulsars: general --- X-rays
}

\def\abstracttext{
We measured temporal and emission properties of quiescent magnetars
using archival Chandra and XMM-Newton data,
produced a list of the properties for 17 magnetars,
and revisited previously suggested correlations between the properties. Our studies carried out with
a larger sample, better spectral characterizations, and more thorough analyses
not only confirmed previously-suggested correlations but also found new ones.
The observed correlations differ from those seen in other neutron-star populations but generally accord with magnetar models.
Specifically, the trends of the intriguing correlations of
blackbody luminosity ($L_{\rm BB}$) with the spin-inferred
dipole magnetic field strength ($B_{\rm S}$) and characteristic age ($\tau_{\rm c}$)
were measured to be $L_{\rm BB}\propto B_{\rm S}^{1.5}$ and $L_{\rm BB}\propto \tau_{\rm c}^{-0.6}$,
supporting the twisted magnetosphere and magnetothermal evolution models for magnetars. We report the analysis results and discuss our findings in the context of magnetar models.
}

\begin{document}
\jkashead %% set title, authors, abstract, etc.

%%%%%%%%%%%%%%%%%%%%%%%%%%%%%%%%%%%%%%%%%%%%%%%%%%%%%%%%%%%%%%%%%%%%%
%%% BEGIN MAIN TEXT HERE %%%%%%%%%%%%%%%%%%%%%%%%%%%%%%%%%%%%%%%%%%%%
%%%%%%%%%%%%%%%%%%%%%%%%%%%%%%%%%%%%%%%%%%%%%%%%%%%%%%%%%%%%%%%%%%%%%

\section{Introduction}
\label{sec:1}
Magnetars are neutron stars with an ultrastrong magnetic field ($B$),
greater than the quantum critical threshold of $4.414\times 10^{13}$\,G.
They have relatively long spin periods ($P\sim 2$--$12$\,s) and large spin-down rates
($\dot P\sim 10^{-13}$--$10^{-10}$\,s\ $\rm s^{-1}$) compared to rotation-powered pulsars (RPPs).
Thus magnetars have strong spin-inferred dipole magnetic-field strengths
$B_{\rm S}=3.2\times 10^{19}\sqrt{P \dot P}\rm\ G$  of typically $>10^{14}$\,G.
Magnetars' emission is mostly in the X-ray band and is thought to be produced
primarily by the decay of internal $B$ \citep{Thompson1995, Thompson1996, Thompson2002}.
Thus, $B$ is the key to understand observational properties of magnetars (e.g., as compared to RPPs).
They are a source of fast radio bursts (FRBs) \citep[e.g.,][]{Chime2020, Bochenek2020}
and can be an important contributor to the gravitational wave (GW) background
\citep[e.g.,][]{Regimbau2006, Ho2016}.
Hence it is very important to characterize and understand magnetars' properties well,
especially in the advent of sensitive instruments for GW \citep[][]{LIGO2015} and
FRB detection \citep[][]{CHIME2018}.

The most distinctive features of magnetars' emission in comparison to RPPs' are bursts and outbursts \citep[e.g.,][]{Kaspi2003, CotiZelati2018}
which occur in magnetars much more frequently than in RPPs \citep[e.g.,][]{Perna2011}.
The internal decay of $B$ can be generated via ambipolar diffusion and
Hall drift \citep[e.g.,][]{Goldreich1992}; these can apply stress to the stellar crust
which may then be sheared. The crust displacement by the magnetic shear may
twist external $B$ and induce various instabilities in the magnetosphere
\citep[e.g.,][]{Beloborodov2009}. They generate diverse transient
behaviors \citep[e.g.,][]{Kaspi2017}, e.g., bursts and outbursts.
Magnetar bursts on a timescale of seconds may be produced by
rapid magnetospheric reconnection in the magnetosphere \citep[e.g.,][]{Lyutikov2003},
and outbursts are suggested to be caused by changes
in the global $B$ structure \citep[e.g.,][]{Beloborodov2007}. These transient phenomena have been used to test
magnetar models and helped to refine them.

Quiescent emissions of magnetars reflect their
persistent properties and can also provide important clues to magnetar physics \citep[e.g.,][]{Kaspi2010}.
Quiescent X-ray spectra of magnetars are well characterized with thermal blackbody
radiation from the surface and/or nonthermal radiation in the magnetosphere,
and modulate on the spin periods. These observed emission properties are similar to
those of other X-ray pulsars, but there are differences; (1) surface temperatures of
magnetars are inferred to be higher than those of other X-ray pulsars, and (2) the spectral
energy distributions of the nonthermal emissions from some magnetars show
a dramatic turn-over at $\sim$10\,keV \citep[][]{Kuiper2006}. Both of them are thought
to be caused by the strong $B$ of magnetars. (1) can be explained as due to the power
supplied by the internal $B$ decay, and (2) may imply that the
magnetospheric emission mechanism of magnetars is different from that of RPPs;
magnetospheric emission of magnetars is thought to be produced by resonant
cyclotron scattering (RCS)
of the thermal photons by electrons in the twisted $B$ field
\citep[e.g.,][]{Thompson2002, Lyutikov2006, Pons2007, Beloborodov2013,Wadiasingh2018}
not by their synchrotron radiation as has been suggested
for RPPs \citep[e.g.,][]{Wang2013,Torres2019}.

Although $B$ is the key to understand the quiescent emission of magnetars,
there may be other important factors. Spin-down power $\dot E_{\rm SD}$ ($\propto \dot P/P^3$)
may play a role for the emission, and spin-down
torque ($\propto \dot \nu$) may be influenced by plasma surrounding the star.
The emission strength and
the physical parameters co-evolve on a timescale of $\sim$Myr via a long-term decay of
$B$ due to conduction and diffusion \citep[e.g.,][]{Pons2007,Vigano2013},
and so the age ($t_{\rm age}$) may also be an important factor.
Impacts of these physical properties ($B$, $t_{\rm age}$ etc.) on the emission
would be manifested by correlation between the physical and radiative properties; such correlations have been seen in populations of X-ray pulsars \citep[e.g.,][]{Li2008,Zhu2011}

Statistical studies of quiescent properties of magnetars have been performed previously.
\citet{Marsden2001} reported a correlation between spin frequency derivative ($\dot \nu$; spin-down torque)
and soft X-ray (0.5--10\,keV) photon index ($\Gamma$).
\citet{Kaspi2010} used a larger sample and suggested that both soft-band ($<10$\,keV) and hard-band ($>10$\,keV)
spectral indices are correlated with $B_{\rm S}$. \citet{Enoto2010}
found that spectral hardness (1--60\,keV flux ratio of a hard and a soft spectral component)
 is correlated with the characteristic age ($\tau_c$) and $B_{\rm S}$.
These helped to develop the twisted-$B$ model
\citep[e.g.,][]{Thompson2002, Beloborodov2013}.
\citet{An2012} found a correlation between $B_{\rm S}$ and 2--10\,keV
X-ray luminosity ($L_{\rm 2-10\,keV}$) using a sample of high-$B$ pulsars and magnetars, and suggested that they share similar
physical processes which are controlled primarily by $B$;
this is in accord with the discoveries of low-$B_{\rm S}$ ($\le 10^{13}$\,G)
magnetars \citep[][]{Rea2013} and magnetar-like outbursts
in typical RPPs \citep[e.g.,][]{Gavriil2008,Archibald2018}.
\citet{Mong2018} confirmed the $B_{\rm S}$-$L_{\rm 2-10\,keV}$ correlation. They
further employed a two-blackbody (2BB) or two-blackbody plus power-law (2BB+PL)
model for the spectra,
and found that the cold BB temperature ($kT_1)$ is correlated with $B_{\rm S}$, lending supports to the long-term magnetothermal evolution models \citep[e.g.,][]{Pons2007,Perna2011,Vigano2013}.
\citet{CotiZelati2018} compiled outburst and quiescent fluxes of magnetars, and performed
a systematic study with emphasis on the outburst properties.

More magnetars have been discovered and their quiescent properties
have been better measured since the previous correlation studies.
Thus, it is timely to update the correlation results with a larger sample, refined
measurements, and thorough analyses.
In this paper, we carefully identified quiescent states of magnetars using information
collected from literature, and selected 17 magnetars whose quiescent properties
could be well characterized (Section~\ref{sec2}).
We measured their emission properties and
investigated correlations between various radiative and temporal properties
including pulsed fractions ($\eta$)
(see Section~\ref{sec:3_2} for the definition of $\eta$
which has not been investigated previously (Sections~\ref{sec:3} and \ref{sec:4}).
We discussed the correlation results in Section~\ref{sec:5},
and summarized them in Section~\ref{sec:6}.

\section{Target Selection and Data Reduction}
\label{sec2}
We selected targets for our study based on the McGill online magnetar
catalog\footnote{http://www.physics.mcgill.ca/$\sim$pulsar/magnetar/main.html}
\citep[][]{Olausen2014} and the magnetar outburst online
catalog \citep[][]{CotiZelati2018}.\footnote{http://magnetars.ice.csic.es/\#/welcome}
The former lists basic temporal and spectral properties of quiescent magnetars,
and the latter provides long-term light curves and helps to identify the epoch of
the faintest state for some magnetars. We made a further literature search to
choose adequate observational data collected in quiescence, and
analyzed them to measure the quiescent properties.
While the previous measurements were useful, we reanalyzed the quiescent data
because some of quantities we intended to investigate
(e.g., BB radius, $\eta$ and PL flux) 
could not be retrieved from the catalogs or literature.
Note that the quiescent data used in this study were taken during one quiescent period, i.e., we did not combine data acquired before and after an outburst.

The targets and data used in this work are listed in Table~\ref{ta:ta1}.
Note that the list is slightly different from that in the McGill
or the outburst catalog because we omitted magnetars whose
quiescent properties could not be measured.
For most of the targets, the quiescent state was well identified by a period with stable and low flux.
But low-cadence observations for some magnetars did not allow a firm identification of the
`stable' quiescent state, and for them we assumed
that the lowest-flux data far away ($>$a few years) from
their outbursts represent well the quiescent states.

We downloaded archival Chandra and XMM-Newton (XMM hereafter) data from the HEASARC data archive and reduced
them following the standard procedures. We reprocessed the Chandra data using
the {\tt chandra\_repro} tool of CIAO~4.13 along with CALDB~4.9.4.
The XMM-Newton data were reduced with the {\tt emproc} and {\tt epproc} tasks of
XMM science analysis system (SAS) v2019, and we filtered out particle flares
from the data following the flare-removal procedure.\footnote{https://www.cosmos.esa.int/web/xmm-newton/sas-thread-epic-filterbackground}
Note that there are Swift data available for some targets, but given the
small effective area of the instrument and short exposures for the observations, the Swift data are less useful.  Note also that Suzaku and NuSTAR observed a few targets, which can help our investigation. We discussed these measurements (Section~\ref{sec:3_3}) but did not use them in this study. We defer further Suzaku and NuSTAR investigations to future work.

\setlength\tabcolsep{1pt}
%\begin{landscape}
\begin{tiny}
%%% Table %%%%%%%%%%%%%%%%%%%%%%%%%%%%%%%%%%%%%%%%%%%%%%%%%%%%%%%%%%%%%%%%%%%%%%%
%\begin{table*}
\begin{landscape}
\begin{table}
\centering
\caption{Spectral and temporal properties of the selected targets}
%\makebox[\textwidth]{
\begin{tabular}{lcllcccccccccccccc}
%%%\toprule
\hline
\#$^{\rm a}$ & Obs ID & Inst.$^{\rm b}$ & Model  & $N_{\rm H}$ & $kT_1$ & ${R_{1,\rm BB}}$ & $kT_2$ & ${R_{2,\rm BB}}$ & $\Gamma$ & ${L_{\rm PL}}$ & $P^{\rm c}$ & $\dot P^{\rm c}$ & $\eta$ & $\xi$ & $d^{\rm c}$ & $\chi^{2}$/dof & Net counts \\
 &        &  & & ($10^{22}\rm \ cm^{-2}$) & (keV)     & (km)          & (keV)             & (km)   &        & (${10^{35}\rm \ erg\ s^{-1}})$ & (s) & ($10^{-11}$)  & & & (10\,kpc) & & \\ \hline
1  & 0304250401  & X  & BB+PL & $0.091^{+0.079}_{-0.038}$ & 0.34(1)  & 9.6(6)  & $\cdots$ &  $\cdots$ & 1.9(1) & 0.7(1) & 8.02  & 1.88   & 0.27(4) & 2  & 6.24 & 109/97 & 3227(58)  \\ 
2  & 0112781101 &  X  & 2BB+PL & 0.70(2) & 0.30(1)  & 13.8(6) & 0.55(2)  &  3.0(3)  & 2.90(4) &  1.4(1)  & 8.69  & 0.20   & 0.07(1) & 2 & 0.36  & 140/131& 163440(405)  \\ 
3  & 0693100101  & X  & BB  & 0.115(6)  & 0.36(2)  & 0.06(1) & $\cdots$ & $\cdots$ & $\cdots$& $\cdots$ & 9.08  & 0.0004 & 0.65(9) & 1 & 0.2 & 33/23 & 486(32)  \\ 
4  & 14811,15564  & C  & BB+PL  & $1.4(1)$  & 0.74(4)  & 0.10(2) & $\cdots$ & $\cdots$ & 3.9(1)  & 0.058(2)  & 5.76  & 0.59   & 0.46(2) & 3 & 0.2  & 261/280 & 8027(90)  \\ 
5  & 10806 & C  & BB+PL & $0.06/0.54(6)$$^{\rm d}$ & 0.45(3)  & 7(1)  & $\cdots$ & $\cdots$ & 2.4(1)  & 5.6(4)   & 8.05  & 3.8    & $\cdots$ & $\cdots$ & 5.36  & 123/140 & 4200(66) \\ 
6 & 6736$^{\rm e}$ & C  & BB+PL  & 0.97(1)  & 0.57(1)  & 2.02(8) & $\cdots$ & $\cdots$ & 3.0(1)  & 1.38(4) & 6.46  & 2.25   & 0.68(1) & 2 & 0.9 & 303/283 & 19768(141)  \\ 
7  & 0402910101  &  X  & BB+PL  & 3.46(3) & 0.40(2)  & 0.8(2)  & $\cdots$ & $\cdots$ & 4.0(1)  & $0.13^{+0.03}_{-0.02}$  & 2.07  & 4.77   & $\cdots$ & $\cdots$ & 0.45 & 140/131 & 4632(70) \\ 
8  & 0742650101  &  X & PL   & 10(2) & $\cdots$ & $\cdots$& $\cdots$ & $\cdots$ & 2.1(3)  & 0.02(1)  & 2.59  & 1.9    & $\cdots$ & $\cdots$ & 1.1 & 53/51 & 175(19) \\ 
9	  & 0404340101  &  X & BB+PL & 2.39(5) & 0.56(5) & 0.17(3) & $\cdots$ & $\cdots$ & 3.8(2) & $0.039^{+0.006}_{-0.005}$  & 10.61 & 0.02   & 0.80(3) & 1  & 0.39 & 129/136 & 2829(57)  \\ 
10   & 4605$^{\rm e}$  &  C  & BB+PL & 1.36(4) & 0.455(4) & 3.3(1)  & $\cdots$ & $\cdots$ & 2.50(2) & 1.42(3)  & 11.01 & 1.95   & 0.324(5) & 3 &  0.38  & 479/440 & 120681(348) \\ 
11   & 0790870201  &  X  & BB+PL & 3.6(2) & 0.61(1)  & 1.56(5) & $\cdots$ & $\cdots$ & 0.9(4)  & $0.12^{+0.03}_{-0.02}$ & 3.83  & 6.4    & 0.29(3) & 1 &  1.32  &234/203 & 8444(98)\\ 
12 & 0654230401  &  X & BB+PL & 9.7(1) & 0.59(3)  & 1.2(1)  & $\cdots$ & $\cdots$ & 1.4(1)  & 0.8(1)   & 7.55  & 49.5   & 0.06(2) & 1 & 0.87  & 241/212 & 14252(121) \\ 
13   & 15870  &  C & 2BB  & 0.92(2)  & 0.17(1)  & 10(1)  & 0.33(2) &  0.9(2)  & $\cdots$ & $\cdots$ & 5.54  & 0.78   & 0.33(3) & 1  & 0.35 & 71/89 & 3095(56) \\ 
14   &  15589--15593 &  C  & BB+PL & 0.453(8) & 0.12(1) & 4(1)  & $\cdots$ & $\cdots$ & 4.5(4)  & $2^{+0.9}_{-0.6}\times 10^{-3}$ & 8.44  & 0.002  & 0.38(3) & 1 & 0.16  & 128/114 & 2157(46) \\ 
15  &   0013340201  & X  & BB+PL & 2.26(5) & 0.41(1)  & 6.1(4)  & $\cdots$ & $\cdots$ & 2.25(6) & 3.3(2)   & 11.79 & 4.09   & 0.15(2)  & 2 & 0.85 & 299/321 & 26306(168)\\ 
16  & 0790610101 &  X & BB+PL & 1.9(1) & 0.52(1)  & 2.6(1)  & $\cdots$ & $\cdots$ & 1.3(2)  & 0.38(4)  & 5.20  & 9.20   & 0.20(2) & 2 & 1.25 & 278/237 & 12053(111)  \\ 
17   & 0057540101  &  X & BB+PL & 1.10(2) & 0.48(2)  & 1.5(1) & $\cdots$ & $\cdots$ & 4.03(5) & 2.4(1)  & 6.98  & 0.05   & 0.18(1) & 2 & 0.32  & 452/447 & 19715(141) \\  \hline

\end{tabular}

\tabnote{
%\footnotesize{
Note. Numbers in parentheses are a quadrature sum of the statistical and systematic (due to $N_{\rm H}$ uncertainties) uncertainties at the 1$\sigma$ level.\\
$^{\rm a}$1. CXOU~J010043.1$-$721134, 2. 4U~0142+61, 3. SGR~0418+5729, 4. SGR~0501+4516, 5. SGR~0526$-$66, 6. 1E~1048.1$-$5937, 7. 1E~1547.0$-$5408, 8. SGR~1627$-$41, 9. CXOU~J164710.2$-$455216, 10. 1RXS~J170849.0$-$400910, 11. CXOU~J171405.7$-$381031, 12. SGR~1806$-$20, 13. XTE~J1810$-$197, 14. Swift~J1822.3$-$1606, 15. 1E~1841$-$045, 16. SGR~1900+14, 17. 1E~2259+586
$^{\rm b}$ Instrument. X: XMM and C: Chandra.\\
$^{\rm c}$ Values taken from the McGill online catalog.\\
$^{\rm d}$ Absorption column densities in the Galaxy and large Magellanic cloud \citep[][]{Park2012}.\\
$^{\rm e}$ Continuous clocking mode observation.\\
\label{ta:ta1}}
%\end{table*}

\end{table}
\end{landscape}
%%%%%%%%%%%%%%%%%%%%%%%%%%%%%%%%%%%%%%%%%%%%%%%%%%%%%%%%%%%%%%%%%%%%%%% % fitting error
\end{tiny}
%\end{landscape}

\section{Data analysis Result}
\label{sec:3}
We describe general analysis procedure first (Sections~\ref{sec:3_1} and \ref{sec:3_2}) and present the results for each target in Section~\ref{sec:3_3}.

\subsection{Spectral analysis}\label{sec:3_1}
Quiescent fluxes of magnetars substantially differ from source to source, and so
we selected adequate source and background regions for each source for spectral analysis.

%%% FIGURE %%%%%%%%%%%%%%%%%%%%%%%%%%%%%%%%%%%%%%%%%%%%%%%%%%%%%%%%%%%%%%%%%%%%
\begin{figure*}
\centering
\includegraphics[width=160mm]{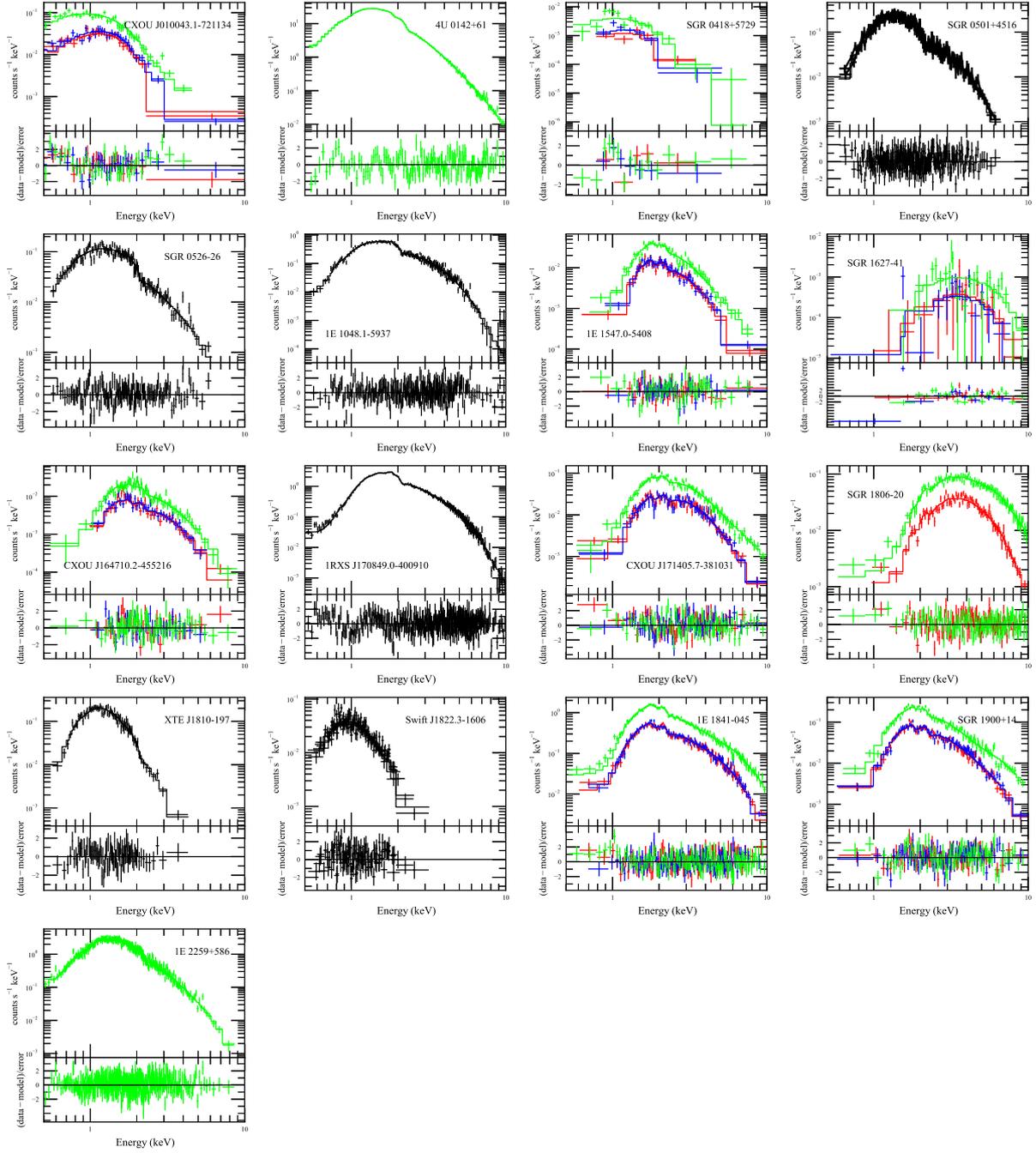}
\vspace{-2mm}
\caption{0.5--10\,keV quiescent spectra of the targets in Table~1. Chandra-measured
spectra are plotted in black, and XMM-measured ones are shown in red (Mos1), blue (Mos2),
and green (PN). The bottom panel in each figure shows residual after subtracting the
best-fit model (solid lines).\label{fig:fig1}}
\end{figure*}
%%%%%%%%%%%%%%%%%%%%%%%%%%%%%%%%%%%%%%%%%%%%%%%%%%%%%%%%%%%%%%%%%%%%%%%%%%%%%%%
For the Chandra data, we used a $R=1''$--$4.5''$ circular region centered at the source position to extract the source spectrum, and used a $R=2''$--$10''$ circle in a nearby source-free region to extract the background spectrum.
Note that we used $5''\times 50''$ and $10''\times 50''$ rectangular regions
for the source and background extraction, respectively, 
for the bright magnetars 1E~1048.1$-$5937 and 1RXS~J170849.0$-$400910 because the
observations were taken with the continuous clocking mode (Table~\ref{ta:ta1}).
For the XMM data analysis, we used a $R=16$--$20''$
and a $R=40$--$60''$ circle for the source and background extraction, respectively.
The target 1E~2259+586 was observed far off-axis, and thus we used a 20$''\times$40$''$
elliptical region for the source spectrum.
We verified that the results did not
alter significantly depending on the source/background region selections.
Corresponding response files were generated with the standard CIAO and SAS tools
for the Chandra and XMM data, respectively. We fit the spectra  employing the $\chi^2$ statistic after
grouping them to have a minimum of 20 events per spectral bin
except for SGR~1627$-$41. For it, we grouped the spectra to have
at least 5 events per bin due to a lack of photon statistics
and utilized the $l$ statistic\footnote{https://heasarc.gsfc.nasa.gov/xanadu/xspec/manual/XSappe\\ndixStatistics.html}
\citep[][]{Loredo1992} for the spectral fit.

Actual emission spectra of magnetars can be very complex to be fully described by simple phenomenological models because the thermal emission from the surface is modified in the atmosphere and magnetosphere \citep[e.g.,][]{Ho2001,Potekhin2001} and the nonthermal emission depends on the poorly known RCS geometry \citep[e.g.,][]{Wadiasingh2018}. In general, a BB, PL, or BB+PL model
has been employed in literature because they can approximately represent the thermal
and/or nonthermal emissions of magnetars. We adopted the BB+PL model as the default since it has been most commonly used in previous studies (Section~\ref{sec:3_3}). We checked to see if the model adequately describes our data, and made a literature search to see if a different model has been favored on statistical or physical grounds. The BB+PL model adequately described the spectra of most of the targets in the previous and our studies (Table~\ref{ta:ta1}). A different model, BB, PL, 2BB, or 2BB+PL, was favored for some targets (e.g., 4U~0142+61, SGR~0418+5729, SGR~1627$-$41, and XTE~J1810$-$197; Table~\ref{ta:ta1}) in previous studies, and our investigation of their spectra agreed with the previous results; for these targets we adopted the favored models. Note also that multiple spectral models (mostly BB+PL and 2BB) could not be discerned for some magnetars. In these cases, we took the BB+PL model as our baseline and investigated the other model as an alternative in the correlation study (see Section~\ref{sec:4_2}).

We fit the source spectra with the aforementioned models in {\tt XSPEC} v12.11.1 \citep[][]{Arnaud1996} and show them in Figure~\ref{fig:fig1}.
Because the absorption column densities ($N_{\rm H}$) towards some targets
were often not well constrained with the quiescent data alone due to the paucity of counts,
we searched literature for $N_{\rm H}$ values that were inferred from a multi-epoch spectral analysis
conducted with the same spectral model as ours.
If such a value was available, we used the same abundance and cross section as those in the literature
and held $N_{\rm H}$ fixed at the value (Section~\ref{sec:3_3}). 
If we could not find a reported value of $N_{\rm H}$, we optimized it in our spectral fit using the {\tt tbabs} model
with {\tt vern} cross section \citep[][]{Verner96} and {\tt angr} abundance \citep[][]{Anders1989}.
The measured spectra and the best-fit parameters are presented in Figure~\ref{fig:fig1} and Table~\ref{ta:ta1}, respectively. The models describe the data adequately (null hypothesis probabilities $>0.03$) although some residuals are noticeable. They may indicate some other physical processes \citep[e.g., cyclotron absorption;][]{Tiengo2013} but an additional component was not required in the spectral fits. 
We describe the analysis results in more detail in Section~\ref{sec:3_3}.

\subsection{Timing analysis}\label{sec:3_2}
%%% FIGURE %%%%%%%%%%%%%%%%%%%%%%%%%%%%%%%%%%%%%%%%%%%%%%%%%%%%%%%%%%%%%%%%%%%%
\begin{figure*}
\centering
\includegraphics[width=170mm]{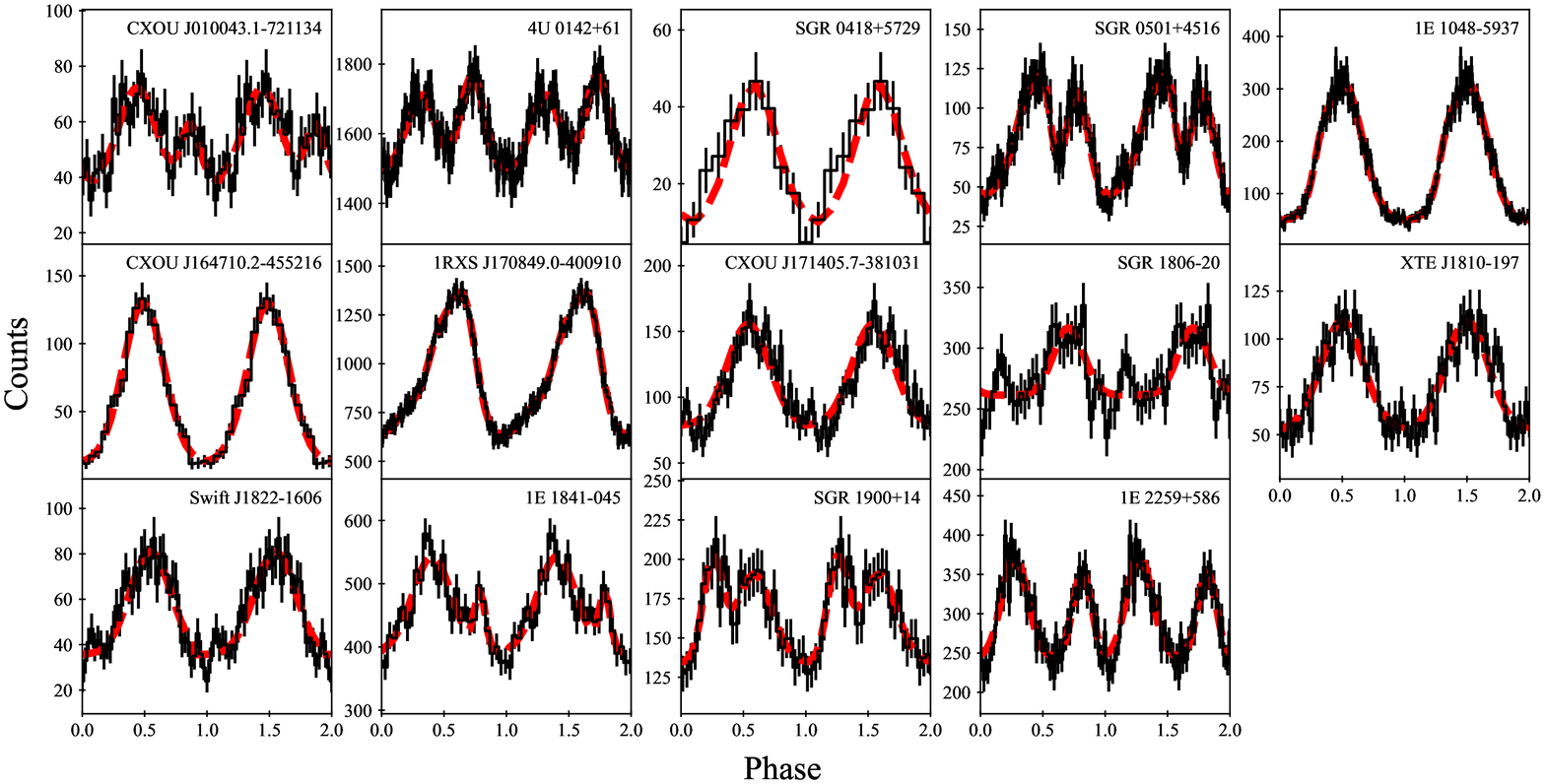}
\vspace{-0mm}
\caption{Background-subtracted 0.5--10\,keV pulse profiles (black) of the targets of which pulsations were significantly detected. The best-fit models are displayed by red dashed lines.
\label{fig:fig2}}
\end{figure*}
%%%%%%%%%%%%%%%%%%%%%%%%%%%%%%%%%%%%%%%%%%%%%%%%%%%%%%%%%%%%%%%%%%%%%%%%%%%%%%%

In our measurements of the temporal properties of the quiescent magnetars,
we attempted to detect their pulsations and measure $P$ at the epoch.
Even though they were already well measured (e.g., McGill magnetar catalog),
our reanalyses were necessary because $P$ changes with time (e.g., slowly due to $\dot P$) and would
be different at the epoch of the observational data we analyzed.
Small inaccuracies ($< 10^{-2}$\,s) in $P$ are not a problem for our correlation study,
but the measurements of $\eta$ may be substantially affected
by the inaccuracies.

For a timing analysis, we barycenter-corrected the source and background event arrival
times using the source positions reported in the McGill magnetar catalog and
searched for pulsations of each target by employing
an $H$ test \citep[e.g.,][]{deJager1989} to measure $P$ of the target at the epoch of the observation.
The pulsations were well detected for most of the targets except
for a few faint sources. The measured $P$ values are not very different ($\Delta P/P\le 0.005$)
from those reported in the McGill catalog.
We then constructed pulse profiles of the source and background emissions by folding 0.5--10\,keV events on the best period.
The background-subtracted pulse profiles are displayed in Figure~\ref{fig:fig2}.

While $\eta$ of a magnetar can be defined in various ways \citep[e.g.,][]{Vogel2014,An2015}, in this work we defined it to be the area fraction above the constant level in the pulse profiles which were modeled as a combination of Gaussian functions.
We fit the profiles of the targets with a Gaussian plus constant function employing the $\chi^2$ statistic and successively added more Gaussians
until an additional Gaussian was unnecessary. 
For each addition of a Gaussian, we performed an $F$ test. Note that the ideal $F$ distribution may be inadequate in the case that the parameter values of the additional component are near their boundaries. Hence we carried out simulations \citep[e.g.,][]{Protassov2002}; we generated 10,000 fake profiles based on the simpler model, fit them with the more complex model to compute the $F$ values, and derived an $F$ distribution appropriate for the model comparison. We required the $p$ value corresponding to the measured $F$ to be less than 0.01 for an addition of a new component. We verified that the optimized models achieved good fits to the pulse profiles with the $\chi^2$ probabilities $>0.2$. The numbers of Gaussians ($\xi$) needed to fit the profile and the estimated $\eta$ are presented in Table~\ref{ta:ta1}. We further verified the $\xi$ values using the unbinned likelihood fit and Akaike information criterion \citep[AIC;][]{aic74}. We found that the $\xi$ values inferred by this method generally agreed with the above results, but in a few cases the results of the AIC and $F$ test differed by 1. Note, however, that the change of $\xi$ by 1 did not have large influence on the $\eta$ estimation (e.g., $\Delta \eta < 1$\% for $\Delta \xi=1$) as long as the model describes the profile adequately; we included this small variation as the systematic uncertainty.

The best-fit functions are displayed in Figure~\ref{fig:fig2}.
Note that the estimated $\eta$ value varies depending weakly on the number of bins. To mitigate this variation,
we changed the number of bins by $\sim 10$\%, regarded the standard deviation of $\eta$ as
a systematic uncertainty and added it to the statistical one. These systematic uncertainties
are comparable to the statistical ones.

Note that we could not determine $\dot P$ for any of the targets with our timing analysis
because the observational data we analyzed covered only a short time interval (e.g., tens of ks)
and thus the effect of $\dot P$ was indiscernible. For this reason,
we use $P$ and $\dot P$ values reported in the McGill magnetar catalog
for the correlation study (see Section~\ref{sec:4_1}).

\subsection{Results of the spectral and timing analyses}
\label{sec:3_3}
Here we briefly describe the spectral and temporal properties of the targets and present our measurements. Note that the errors on the parameters reported in this section are 1$\sigma$ uncertainties obtained from spectral fitting (with fixed $N_{\rm H}$ unless noted otherwise).

{\noindent \bf CXOU~J010043.1$-$721134}
in the small Magellanic cloud is a persistent magnetar whose emission has been stable
for approximately three decades \citep[][]{McGarry2005,Chatterjee2021}.
A previous study \citep[][]{Tiengo2008} of the source performed with multi-epoch data suggested
that its quiescent spectrum is best described by a 2BB model with
$N_{\rm H}=(6.3^{+2.0}_{-1.6})\times 10^{20}\rm \ cm^{-2}$, $kT_1=0.30\pm0.02$\,keV
$kT_2=0.68^{+0.09}_{-0.07}$\,keV while other commonly used spectral models (e.g., BB+PL)
were disfavored by the multi-epoch data.
This magnetar's $\eta$ was measured to be $32\pm3$\% in the 0.2--6\,keV \citep[][]{Tiengo2008}.

We used the longest XMM data acquired on 2005 November 27, measured the quiescent spectrum,
and fit it with a BB+PL and a 2BB model. It is beneficial to use low-energy (e.g., $<$0.5\,keV) data for this source with very low $N_{\rm H}$, and hence we fit the data in the 0.1--10\,keV band as was done by \citet{Tiengo2008}. We found that both spectral models adequately describe the spectrum, and the best-fit parameters are
$kT_1=0.34\pm0.01$\,keV and $\Gamma=1.9\pm0.1$
\citep[for $N_{\rm H}=9.1\times 10^{20}\rm \ cm^{-2}$;][]{Tiengo2008} for
the BB+PL model,
and $kT_1=0.29\pm0.02$\,keV and $kT_2=0.57\pm0.07$\,keV 
(for $N_{\rm H}=6.3\times 10^{20}\rm \ cm^{-2}$)
for the 2BB model. For this source, we use the BB+PL parameters as our baseline
for the correlation study (Section~\ref{sec:4_1}), but
we also consider the 2BB parameters (Section~\ref{sec:4_2}). Note that $kT_2$ of the 2BB model
is slightly different (but within the uncertainty) from the previously reported value;
this seems to be caused by the covariance between the BB temperature and radius.
Our fit favors a smaller $kT_2$ and a larger $R_{\rm 2, BB}$.
We verified that this difference did not have significant influence on the correlation results. We measured $\eta$ to be $27\pm4$\% in the 0.5--10\,keV band.

{\noindent \bf 4U~0142+61} is
an old ($\tau_c=68$\,kyr), bright, and relatively stable magnetar \citep[][]{Rea2007a,Tendulkar2015},
albeit with small and infrequent activities \citep[e.g.,][]{Gogus2017,Archibald2017,CotiZelati2018}.
\citet{Gonzalez2010} analyzed multiple observations spread over 2000\,d and
suggested that the source's quiescent spectrum is mildly variable and is best described by 2BB+PL.
The authors reported ranges of the spectral parameters:
$kT_1=$0.27--0.31\,keV, $kT_2=$0.50--0.60\,keV, and
$\Gamma=$2.6--3.0 for $N_{\rm H}=(7.0\pm0.2)\times 10^{21}\rm \ cm^2$.

We selected the observation made with XMM on 2003 January 24 when the source was faintest. Because the MOS data were taken with the timing mode, we used only the PN data.
Our 2BB+PL fit of the spectrum resulted in $kT_1=0.30\pm0.01$\,keV, $kT_2=0.55\pm0.02$\,keV,
and $\Gamma=2.90\pm0.04$, which are consistent with the previous measurements.
The simpler BB+PL and 2BB models could be rejected statistically.
The magnetar's $\eta$ is known to be small (Fig.~\ref{fig:fig2}),
and we found $\eta=7.4\pm0.5$\% in the 0.5--10\,keV band, which is consistent with
a previous result of $7.7\pm0.9$\% in the 0.3--10\,keV band \citep[e.g.,][]{Gohler2005}.

{\noindent \bf SGR~0418+5729} is a
low-$B$ ($B_{\rm S}=6\times 10^{12}$\,G) magnetar that was discovered due to its dramatic
outburst on 2009 June 5 \citep[][]{Rea2013}.
The source flux had declined during the subsequent
years and reached a quiescent level $\sim$2\,yr after the onset of the outburst.
\citet{Rea2013} analyzed long-term data spanning 3\,yrs, fit
the quiescent spectrum with a simple BB model having
$N_{\rm H}=(1.15\pm0.06)\times 10^{21}\rm \ cm^{-2}$ and $kT=0.32\pm0.05$\,keV,
and measured $\eta$ to be $\sim$60--80\% in the 0.5--10\,keV band.

We analyzed the most sensitive XMM data taken on 2012 August 25. The source spectrum was well fit with a BB model having $kT_1=0.36\pm0.02$\,keV, and additional spectral components were statistically unnecessary. The source's pulsations with $\eta=65\pm9$\% were well detected.

{\noindent \bf SGR~0501+4516}
underwent an outburst in 2008 (MJD 54701; 2008 August 23) and its flux decayed to a flat level
$\sim$1\,yr after the outburst \citep[][]{Camero2014}.
To measure the quiescent emission properties,
\citet{Rea2009} analyzed archival ROSAT data collected
between 1992 September 21 and 24 (before the 2008 outburst),
and measured the 0.1--2.4\,keV spectrum to be a BB ($kT=0.38_{-0.15}^{+0.36}$\,keV)
or a PL ($\Gamma>0.6$) for $N_{\rm H}=6^{+5}_{-3}\times 10^{21}\rm \ cm^{-2}$
with the 1--10\,keV observed fluxes of $1.3\times10^{-12}\rm \ erg\ cm^{-2}\ s^{-1}$
and $4.2\times 10^{-12}\rm \ erg\ cm^{-2}\ s^{-1}$ for the BB and PL model, respectively.
On the other hand, \citet{Mong2018} performed a spectral analysis
with multi-epoch data spread over $\sim$5\,yrs (`post-burst' Chandra and Suzaku data taken between 2008 and 2013) and found that
a BB+PL model with $kT_1=0.63^{+0.04}_{-0.05}$\,keV, $\Gamma=3.9^{+0.3}_{-0.2}$
for $N_{\rm H}=(1.43^{+0.09}_{-0.08})\times 10^{22}\rm \ cm^{-2}$, or
a 2BB+PL model with $kT_1=0.26^{+0.01}_{-0.02}$\,keV, $kT_2=0.62^{+0.03}_{-0.04}$\,keV
and $\Gamma=2.3^{+0.7}_{-2.5}$ for $N_{\rm H}=(9.0\pm0.2)\times 10^{21}\rm \ cm^{-2}$
fits the data well; they favored the latter based on some high-energy residuals.

The source fluxes measured by \citet{Mong2018} are higher by $\sim$50\% than
that measured with ROSAT, possibly indicating that the source had not returned
to its quiescent state or that the source's quiescent emission is variable.
We, however, speculated that this discrepancy might be caused by ROSAT's lack
of $>$2\,keV sensitivity.
To confirm this, we used two Chandra observations taken on 2012 December 9 and 2013 April 3 when the source
flux decreased by an order of magnitude compared to the maximum \citep[see][]{Camero2014}.
For a spectral analysis we restricted the energy band to `0.5--2.4\,keV' to compare with the ROSAT results
and held $N_{\rm H}$ fixed at $6\times 10^{21}\rm \ cm^{-2}$ which was used for the ROSAT analysis (see above).
We found out that the `Chandra' spectra are well modeled with
a BB having $kT=0.39\pm0.01$\,keV with the 1--10\,keV observed flux
of $1.4\times10^{-12}\rm \ erg\ cm^{-2}\ s^{-1}$.
These results are in good agreement with the ROSAT measurements,
meaning that the magnetar actually returned to the preburst state at the
epoch of the Chandra observations.

Our analysis of the aforementioned Chandra data in the `0.5--10\,keV' band ruled out
the simple BB model,
and thus we used BB+PL, 2BB and 2BB+PL models \citep[e.g.,][]{Mong2018} and found that the PL component of
the 2BB+PL model is statistically unnecessary. The best-fit parameters for the
BB+PL model are $kT_1=0.74\pm0.04$\,keV and $\Gamma=3.9\pm0.1$, and those for
the 2BB model are $kT_1=0.27\pm0.01$\,keV and $kT_2=0.72\pm0.02$\,keV.
Our best-fit BB temperatures do not agree well with those of \citet{Mong2018}. This discrepancy is presumably induced by parameter covariance; holding $kT_2$ and $\Gamma$ fixed at their values in our fits reproduces their results. Note also that they jointly analyzed the Chandra and Suzaku data (2013 August), whereas we used only the Chandra data.
We verified that this difference in $kT_2$ did not have a significant impact on the correlation results (Section~\ref{sec:4}).
We use our best-fit BB+PL model for the correlation study,
but also consider the 2BB and 2BB+PL models (Section~\ref{sec:4_2}).
We measured $\eta=46\pm2$\% (0.5--10\,keV),
which is consistent with $45\pm6$\%
in the 0.3--12\,keV band measured when the source was slightly brighter
\citep[with 2009 XMM data;][]{Camero2014}.

{\noindent \bf SGR~0526$-$66}
was discovered in the supernova remnant (SNR) N49 in the large Magellanic cloud (LMC).
The source exhibited a giant flare in 1979 \citep[][]{Cline1982} but no strong
activity has been seen since then. \citet{Guver2012} found that
the source flux has decreased slowly by $\sim$20--30\% over 15\,yrs by comparing
1994 ROSAT \citep[][]{Rothschild1994}, 2000/2001 and 2009 Chandra measurements;
no significant change of the BB temperature was seen.
Given the small flux decay rate, we assumed that the source
has been in (near) a quiescent state over the 15-yr period.
A previous spectral analysis of multi-epoch data found that a BB+PL or a 2BB model
adequately explains the observed spectrum \citep[][]{Park2012}.
The best-fit parameters inferred from the analysis are 
$N_{\rm H, Gal}=6\times 10^{20}\rm \ cm^{-2}$ (Galactic absorption),
$N_{\rm H, LMC}=(5.44^{+0.58}_{-0.59})\times 10^{21}\rm \ cm^{-2}$ (LMC absorption),
$kT_1=0.44\pm0.02$\,keV, and $\Gamma=2.5^{+0.11}_{-0.12}$ for the BB+PL model, and
$N_{\rm H, Gal}=6\times 10^{20}\rm \ cm^{-2}$,
$N_{\rm H, LMC}=1.70^{+0.25}_{-0.23}\times 10^{21}\rm \ cm^{-2}$,
$kT_1=0.39\pm0.01$\,keV, and $kT_2=1.01^{+0.11}_{-0.09}$\,keV for the 2BB model.
Note that the authors favored the 2BB model because the fit-inferred $N_{\rm H, LMC}$ towards the magnetar
agrees better with that inferred towards the surrounding SNR.
Pulsations of this magnetar were only weakly detected, and the reported
$\eta$ values are diverse; \citet[][]{Tiengo2009} measured it to be $13.6\pm0.9$\% in
the 0.65--12\,keV band, whereas \citet{Guver2012} reported $\eta$ of 1.5--4.4\%
in the 0.5--6.5\,keV band. The discrepancy might be caused by different definitions
of $\eta$ and energy bands used for the measurements.

We used the Chandra data acquired on 2009 September 19 when the source flux was lowest \citep[see][]{Guver2012}.
For an analysis of the Chandra data,
we adopted the BB+PL and 2BB models and found
$kT_1=0.45\pm0.03$\,keV and $\Gamma=2.4\pm0.1$ (BB+PL), and
$kT_1=0.40\pm0.02$\,keV and $kT_2=1.0\pm0.1$\,keV (2BB),
using the corresponding absorption column densities reported by \citet{Park2012}
for each of the spectral models.
We use the BB+PL parameters for the correlation study, but also consider the
2BB ones (Section~\ref{sec:4_2}).
The source pulsations were not well detected in the Chandra data and so we could not reliably measure $\eta$ for this source.

{\noindent \bf 1E~1048.1$-$5937}
is a bright magnetar that has exhibited frequent outbursts and bursts \citep[e.g.,][]{Gavriil2004,An2014,Archibald2020}. Its quiescent state was well identified by long-term monitoring \citep[e.g.,][]{Tam2008,Archibald2020}, and \citet{Tam2008}
analyzed Chandra monitoring observations taken in a stable and quiescent state. They found that a BB+PL model with $N_{\rm H}=(0.97\pm0.01)\times 10^{22}\rm \ cm^{-2}$, $kT_1=0.50$--$0.56$\,keV,
and $\Gamma=2.72$--$3.14$ describes the source spectra well, and that
$\eta$ is 61--70\% in the 2--10\,keV band \citep[see also][]{Yang2016}.

We analyzed the Chandra data taken on 2006 September 23,
fit the spectrum with a BB+PL model, and found the best-fit parameters
to be $kT_1=0.57\pm0.01$\,keV and $\Gamma=3.0\pm0.1$.
The source pulsations were well detected
and $\eta$ was measured to be $68\pm1$\% in the 0.5--10\,keV band.

{\noindent \bf 1E~1547.0$-$5408}
is a radio magnetar \citep[i.e., radio pulsations;][]{Gelfand2007,Camilo2007}
that underwent an outburst in 2007 \citep[e.g.,][]{Halpern2008}. The source was in the lowest flux
state in 2006 just before the outburst \citep[][]{Bernardini2011}.
The authors analyzed multi-epoch data taken in outburst (in 2009) and in quiescent (in 2006), fit the quiescent spectrum with a BB+PL model, and measured $kT_1=0.43\pm0.03$\,keV, $\Gamma=4.0\pm0.2$, and $N_{\rm H}=(3.46\pm0.03)\times 10^{22}\rm\ cm^{-2}$.
Note that the $N_{\rm H}$ value was inferred from a joint fit of the multiple-epoch data.
In the quiescent data, \citet{Bernardini2011} did not detect significant pulsations
and hence reported an upper limit of $\le$15\% for $\eta$.

We also analyzed the XMM data collected on 2006 August 21 and were able to rule out single component models (BB or PL).
For the BB+PL model, we measured the best-fit parameters to be
$kT_1=0.40\pm0.02$\,keV and $\Gamma=4.0\pm 0.1$.
Our search for the 2-s pulsations resulted in an insignificant detection
in agreement with \citet{Bernardini2011} \citep[see also][]{Gelfand2007}.

{\noindent \bf SGR~1627$-$41}
has shown outbursts in 1998 and 2008 \citep[e.g.,][]{Kouveliotou1998,Kouveliotou2003,Esposito2008}.
Although a `stable' flux state of the source could not be convincingly identified in
a long-term light curve \citep[][]{An2018}, the source flux reached a
historical minimum in 2015, $\ge$2000\,d after the latest outburst;
we assumed that the magnetar was in quiescence at that time.
The source spectrum measured with the 2015 XMM data was modeled by a
$\Gamma=2.0\pm0.3$ PL \citep[][]{An2018} for $N_{\rm H}=(1.0\pm0.2)\times 10^{23}\rm \ cm^{-2}$ \citep[][]{Esposito2008}.

We analyzed the 2015 XMM data and found $\Gamma=2.1\pm0.3$ for $N_{\rm H}=10^{23}\rm \ cm^{-2}$.
A BB model also achieves a good fit with $N_{\rm H}=(7\pm3)\times10^{22}\rm \ cm^{-2}$
and $kT_1=1.2\pm0.1$\,keV, but the temperature appears to be too high compared to those of other quiescent magnetars.
So we do not consider the BB model in our correlation study.
Its pulsations were detected only during an outburst (bright) state \citep[][]{Esposito2008}, and the
2015 XMM data were not sufficient for a detection of the pulsations.
Note that the correlation results did not change much whether or not we included this source.

{\noindent \bf CXOU~J164710.2$-$455216}
is a low-$B$ magnetar possibly associated with the star cluster Westerlund~1
and has a characteristic age of $\sim$Myr \citep[][]{Muno2006}.
It was in a stable low-flux (`quiescent') state with the 2--10\,keV flux of (2--3)$\times10^{-13}\rm \ erg\ cm^{-2}\ s^{-1}$
\citep[][]{An2013a,CotiZelati2018} before MJD~54000 when the source underwent an outburst.
\citet{Muno2007} fit XMM data taken on 2006 September 16 (MJD~53995; a week before the outburst)
with a BB model and inferred $kT_1$ to be $0.54\pm0.01$\,keV (for $N_{\rm H}=1.28\times 10^{22}\rm \ cm^{2}$),
but a joint analysis of long-term data \citep[including the XMM data of][]{Muno2007} suggested that a BB+PL model with
$N_{\rm H}=(2.39\pm0.05)\times 10^{22}\rm  cm^{-2}$, $kT_1=0.59\pm0.06$\,keV
and $\Gamma=3.86\pm0.22$ explains the spectrum better \citep[][]{An2013a}.

We analyzed the XMM data taken on MJD~53995 and found that
a BB+PL model with $kT_1=0.56\pm0.05$\,keV and $\Gamma=3.8\pm0.2$
%and $N_{\rm H}=2.39\times 10^{22}\rm  cm^{-2}$
explains the data adequately.
A simple BB model could be ruled out with an $F$-test probability of $4\times 10^{-9}$.
Although the pulsations of the source were well detected, $\eta$ of the source measured with
the XMM data has not been reported previously. We detected its pulsations with
high significance
and measured the pulsed fraction to be $\eta=80\pm3$\%. Note that $\dot P$ for this magnetar has been controversial \citep[][]{RodriguezCastillo2014} but a recent study carried out with NICER data found $\dot P\approx 2\times 10^{-13}$ \citep[][]{An2019}. We reported this value in Table~\ref{ta:ta2}.

{\noindent \bf 1RXS~J170849.0$-$400910}
is a very bright magnetar whose emission was detected $>$100\,keV \citep[][]{Kuiper2006}.
This magnetar has not exhibited any outburst, and its flux seems mildly ($\sim$50\%) variable \citep[][]{Cacmaz2013,Rea2007b}.
A previous analysis of $\sim$10\,yr data found that the source's
quiescent spectrum is well described by a BB+PL model having
$kT_1$=0.41--0.48\,keV, $\Gamma$=2.5--2.8, and
$N_{\rm H}=(1.36\pm0.04)\times 10^{22}\rm \ cm^{-2}$, and that
$\eta$ is $35.4\pm 0.5$\% in the 0.3--8\,keV band \citep[][]{Rea2007b}.

We used the Chandra observation taken on 2004 July 3 when the source had the
lowest flux. We fit the source spectrum with a BB+PL model and
measured the best-fit parameters to be $kT_1=0.455\pm0.004$\,keV and $\Gamma=2.50\pm0.02$ which
are within the ranges of the previous estimations.
 The spectral fit was acceptable with the null hypothesis probability of 0.1, but some residual trends are noticeable at low energies (Fig.~\ref{fig:fig1}). Note that the observation was taken with the CC mode and thus an accurate estimation of the background was difficult, which might cause the residual.
Its 11-s pulsations were significantly detected and $\eta$ was measured to be $32.4\pm0.5$\% in our analysis.

{\noindent \bf CXOU~J171405.7$-$381031}
is a bright 3.8-s magnetar associated with the SNR CTB~37B \citep[][]{Halpern2010}.
The source has been stable for $\sim$10\,yrs without exhibiting any burst
or outburst since the Chandra discovery in 2010.
\citet{Gotthelf2019}  jointly analyzed XMM /PN and NuSTAR data taken on 2016 September  22--23,
fit the spectrum with a BB+PL model having
$N_{\rm H}=(3.6\pm0.5)\times 10^{22}\rm \ cm^{-2}$, $kT_1=0.62\pm0.04$\,keV and $\Gamma=0.9\pm 0.3$  (90\% confidence interval).
They measured $\eta$ to be $44\pm4$\% in the 1--5\,keV band.

We analyzed the same XMM data  (MOS+PN) and were able to reproduce the previous results:
$kT_1=0.61\pm0.01$\,keV and $\Gamma=0.9\pm0.4$.
The pulsations of the source were detected with high significance, and we
measured $\eta=29\pm3$\% in the 0.5--10\,keV band.
This is slightly lower than the previous measurement of $44\pm4$\%; the
discrepancy may stem from the different energy bands  as this magnetar shows a change of the pulse profile with energy \citep[][]{Gotthelf2019}.

Note that $\Gamma$ we inferred from the XMM data varies substantially (by $\Delta \Gamma=0.4$; Table~\ref{ta:ta1}) when we varied $N_{\rm H}$ within its 68\% uncertainty. This $\Gamma$ uncertainty can be significantly reduced by jointly analyzing NuSTAR hard X-ray data as was done by \citet{Gotthelf2019}. We therefore jointly analyzed  the XMM (PN+MOS) and NuSTAR (Obs. ID 30201031002) observations, and were able to better constrain both $N_{\rm H}$ and $\Gamma$ to within $1.3\times 10^{21}\rm \ cm^{-2}$ and 0.1, respectively. Nonetheless, we use the large uncertainty ($\Delta \Gamma=0.4$) for our simulations (i.e., for $\langle \kappa_{1,2}\rangle$; see Section~\ref{sec:4}) to be conservative. Note that the best-fit parameters do not change in our joint analysis of the XMM and NuSTAR data and hence the correlation results (i.e., $\kappa_{1,2}$; see Section~\ref{sec:4}) do not alter.

{\noindent \bf SGR~1806$-$20}
is one of the few magnetars that have exhibited a giant flare
\citep[2004 December 27.;][]{Hurley2005,Mereghetti2005,Terasawa2005}.
The source flux has been low and stable since 2010, which
establishes well the quiescent state of the source \citep[][]{Younes2017,CotiZelati2018}.
Its quiescent emission (after 2010) was slightly lower than a preburst level
\citep[i.e., before 2004;][]{Younes2015},
and was well characterized by a BB+PL model with $kT_1$=0.59--0.67\,keV,
$\Gamma=$1.27--1.38, and $N_{\rm H}=(10\pm3)\times10^{22}\rm \ cm^{-2}$
\citep[2015--2016 NuSTAR data;][]{Younes2017}.
These are consistent with results of the multi-epoch XMM data analysis 
\citep[][]{Younes2015} which constrained $N_{\rm H}$ better
($(9.7\pm0.1)\times 10^{22}\rm \ cm^{-2}$).
The root-mean-square (RMS) $\eta$ of the source was measured to be 3--8\% over the 8\,yrs
between 2003 and 2011 \citep[][]{Younes2015}.

Because the BB emission can be more accurately measured with XMM than NuSTAR,
we analyzed the 2011 XMM data to characterize the quiescent emission of the source. Note that we did not use the MOS2 data because they were taken with the timing mode.
We fit the XMM spectrum with a BB+PL model and found $kT_1=0.59\pm0.03$ and $\Gamma=1.4\pm0.1$
\citep[for $N_{\rm H}=9.7\times 10^{22}\rm \ cm^{-2}$;][]{Younes2015}.
Although the source is bright at X-rays, the detection significance for its pulsations in the quiescent data was modest
because of the low $\eta$ \citep[e.g.,][]{Woods2007} which we measured to be $6\pm2$\%.

{\noindent \bf XTE~J1810$-$197}
was serendipitously discovered on 2003 July 15
during its relaxation into a quiescent state after an undetected outburst between
2002 November 17 and 2003 January 23 \citep[][]{Ibrahim2004}.
The source flux in 2003 was higher by approximately two orders of magnitude compared
to the historical minimum measured by ROSAT in 1991--1993 \citep[][]{Gotthelf2004}.
Analyses of the ROSAT survey data suggested that the preburst spectrum of the source is well
described with a BB model having $kT_1=0.18\pm0.02$\,keV and the 0.5--10\,keV
absorbed flux of (5.5--8.3)$\times 10^{-13}\rm \ erg\ cm^{-2}\ s^{-1}$
for $N_{\rm H}=6.3\times 10^{21}\rm \ cm^{-2}$ \citep[][]{Gotthelf2004},
or a 2BB model having $kT_1=0.16\pm0.03$\,keV, $R_{1, \rm BB}=16\pm5$\,km,
$kT_2 = 0.26\pm0.06$\,keV and $R_{2, \rm BB} < 5$\,km 
for $N_{\rm H}=(7.5\pm0.8)\times 10^{21}\rm \ cm^{-2}$
\citep[][]{Bernardini2009}. The source flux seemed
to have reached the stationary preburst level
since 2007 \citep[][]{Alford2016,Pintore2019}.
The lowest flux state after the outburst was observed by Chandra
on 2014 March 1 (a decade after the outburst),
and \citet{Vurgun2019} fit the source spectrum with a 2BB model, inferring
$N_{\rm H}=(9.2\pm 0.2)\times 10^{21}\rm \ cm^{-2}$, $kT_1=0.18\pm0.01$\,keV and $kT_2=0.36\pm0.01$\,keV.

We reanalyzed the 2014 Chandra data, measured the source spectrum, and
fit it with a BB+PL or 2BB model.
The best-fit BB+PL parameters are $kT_1=0.32\pm0.03$\,keV,
$\Gamma=7\pm1$ for $N_{\rm H}=(1.7\pm0.2)\times 10^{22}\rm \ cm^{-2}$,
and the 2BB parameters are
$kT_1=0.17\pm0.01$\,keV and $kT_2=0.33\pm0.02$\,keV
\citep[for $N_{\rm H}=9.2\times 10^{21}\rm \ cm^{-2}$;][]{Vurgun2019}.
Note that we optimized $N_{\rm H}$ for the BB+PL model because it has not been explored previously.
The PL index inferred from the BB+PL model is uncomfortably large. Presumably,
it is an artifact caused by forcing to fit high-temperature BB emission with the PL model.
For this reason, the 2BB model has been favored over the BB+PL one
for this source \citep[e.g.,][]{Gotthelf2004, Alford2016}.
We, therefore, use the 2BB parameters for the correlation study (Section~\ref{sec:4_1}).
We estimated $\eta$ to be $33\pm3$\% in the 0.5--10\,keV band,
which is slightly lower than $41\pm5$\% measured in the 1.5--5\,keV
band \citep[][]{Alford2016}.

Note that the 0.5--10\,keV absorbed flux of $8.3\times 10^{-13}\rm \ erg\ cm^{-2}\ s^{-1}$
and $kT_2$ inferred from the 2BB fit are slightly higher
than but within the uncertainties of the ROSAT measurements.
This may indicate some variability. Alternatively, it may mean that the
Chandra data do not represent very well the quiescent state of the source.
Hence we also investigate the correlations with the ROSAT-measured spectral parameters (Section~\ref{sec:4_2}) and verified that the results did not alter significantly.

{\noindent \bf Swift~J1822.3$-$1606}
is a faint low-$B$ magnetar
\citep[][]{Scholz2014,RodriguezCastillo2016}
that was discovered in outburst on 2011 July 14 \citep[][]{Cummings2011,Livingstone2011}.
The source flux has decreased since then and seemed to have reached a stationary
level $\sim$1000\,d after the onset of the outburst \citep[][]{CotiZelati2018}.
A `preburst' spectrum of the source measured with archival ROSAT data collected
in 1993 was fit with a BB model having $kT=0.12\pm0.02$\,keV
and an absorbed 0.1--2.4\,keV flux of $9_{-9}^{+20}\times10^{-14}\rm \ erg\ cm^{-2}\ s^{-1}$
for $N_{\rm H}=(4.53\pm0.08)\times 10^{21}\rm \ cm^{-2}$ \citep[][]{Scholz2012}.
Note that this $N_{\rm H}$ value was inferred from a joint fit of multi-epoch Swift and Chandra data with a BB+PL model.
The ROSAT data were not sensitive enough to allow a detection of the pulsations.
\citet{Scholz2012} therefore used Swift observations taken when the source was slightly brighter than
at the ROSAT epoch and measured $\eta$ to be 45--50\% in the 2--10\,keV band.

To measure the `post-burst' quiescent spectrum, \citet{Mong2018} analyzed five Chandra observations
taken between 2014 April 14 and 2014 October 11,
and reported the best-fit parameters of a 2BB model of
$kT_1=0.11\pm0.01$\,keV and $kT_2=0.29\pm0.03$\,keV for $N_{\rm H}=(6.2\pm0.5)\times 10^{21}\rm \ cm^{-2}$.
We also analyzed the five Chandra observations and
fit the source spectra with a BB+PL or 2BB model.
These models explain the observed spectra equally well, and 
the best-fit parameters for the BB+PL model are $kT=0.12\pm0.01$\,keV and $\Gamma=4.6\pm0.4$
\citep[for $N_{\rm H}=4.53\times 10^{21}\rm \ cm^{-2}$;][]{Scholz2014},
and those for the 2BB model are $kT_1=0.11\pm0.01$\,keV and $kT_2=0.27\pm0.02$\,keV
\citep[for $N_{\rm H}=6.2\times 10^{22}\rm \ cm^{-2}$;][]{Mong2018}.

To compare with the ROSAT results, We tried to fit the 0.1--2.4\,keV Chandra spectrum with a simple BB model.
The Chandra data required an additional model component, and the measured 0.1--2.4\,keV absorbed flux of $1.7\times10^{-13}\rm \ erg\ cm^{-2}\ s^{-1}$
is $\sim$2 times higher than (but within the uncertainty of) the ROSAT measurement.
This may indicate some variability of the quiescent flux or
alternatively suggest that the source did not reach the quiescent state (e.g., ROSAT measured state) in 2014.
Although the large uncertainties in the ROSAT measurements preclude a firm conclusion,
the differences are not large, meaning that the source was in or `near' quiescence.
Thus, we use the Chandra BB+PL parameters as our baseline for the correlation study,
but investigate the correlations with the ROSAT-measured properties as well (Section~\ref{sec:4_2}).
The source's pulsations were well detected in the Chandra data,
and we measured $\eta$ to be $38\pm3$\%.

{\noindent \bf 1E~1841$-$045}
is the power source of the SNR Kes~73 and is the first magnetar from which $>$100\,keV emission was discovered
\citep[][]{Molkov2004}. It has not shown a dramatic outburst,
but hard X-ray bursts have been detected \citep[e.g.,][]{An2015}.
The source emission is very strong and stable in the X-ray band,
and so its quiescent spectrum was relatively well measured;
\citet{Morii2003} and \citet{An2013b} fit the spectrum with a BB+PL model
and found the best-fit parameters to be $kT=0.42$--$0.44$\,keV and $\Gamma=2.0$--$2.1$
for $N_{\rm H}=(2.2$--$2.3)\times 10^{22}\ \rm cm^{-2}$.

We measured the source spectrum using the XMM data taken on 2002 October 7,
fit the spectrum with a BB+PL model, and obtained the best-fit parameters of
$kT=0.41\pm0.01$\,keV and $\Gamma=2.2\pm 0.1$
\citep[for $N_{\rm H}=(2.26\pm 0.05)\times 10^{22}\rm \ cm^{-2}$;][]{An2013b}.
The source's pulsations were well detected,
and the measured $\eta$ is $15\pm2$\%, which is consistent with
$19\pm3$\% in the 0.6--7\,keV band \citep[][]{Morii2003}.

{\noindent \bf SGR~1900+14}
is a magnetar that exhibited a giant flare \citep[1998 August 27;][]{Feroci1999}.
The source flux appears to have reached a
quiescent level since 2006 \citep[][]{Mereghetti2006,Tamba2019}.
\citet{Tamba2019} fit quiescent spectra measured by XMM and NuSTAR with
a BB+PL model having $kT=0.52^{+0.02}_{-0.01}$\,keV,
$\Gamma=1.4\pm0.3$, and $N_{\rm H}=(1.9\pm0.1)\times 10^{22}\rm \ cm^{-2}$,
and measured $\eta$ to be 19--22\% in the 1--10\,keV band.

We analyzed the XMM data taken on 2016 October 20 \citep[the XMM data analyzed by][]{Tamba2019} when the source flux was lowest.
The XMM spectra were well fit with a BB+PL model having $kT=0.52\pm0.01$\,keV
and $\Gamma=1.3\pm0.2$,
and $\eta$ was measured to be $20\pm2$\% in the 0.5--10\,keV band.

{\noindent \bf 1E~2259+586}
is a bright magnetar within the SNR CTB~109, and
exhibited outbursts in 2002 \citep[][]{Kaspi2003} and 2014--2016.
A long-term light curve \citep[][]{Zhu2008} showed that the source flux has declined relatively rapidly
to a quiescence level on a timescale of a few years.
\citet{Woods2004} measured a quiescent spectrum of the magnetar using
the XMM observation taken on 2002 January 22, just before the 2002 outburst,
and reported the best-fit BB+PL parameters of $kT=0.49\pm0.01$\,keV, and
$\Gamma=4.04\pm0.08$ for $N_{\rm H}=(1.10\pm0.02)\times 10^{22}\rm \ cm^{-2}$.
%\HAR{2--10\,keV abs. flux 1.24e-11, unabs flux 1.53e-11}

We analyzed the same XMM data that \citet{Woods2004} used and fit the spectrum with a BB+PL model.
Note that we used only the PN data because the MOS data are severely affected by the pile-up effect \citep[][]{Woods2004}.
The best-fit BB+PL parameters are $kT=0.48\pm0.02$\,keV and $\Gamma=4.03\pm0.05$.
The pulsations of the source were detected with high significance
and $\eta$ was measured to be $18\pm1$\% which is
consistent with a previous measurement of $23\pm5$\%
in the 2--10\,keV band \citep[][]{Woods2004,Zhu2008}.

\section{Correlation analysis}
\label{sec:4}

\subsection{Correlation study with the baseline model parameters}
\label{sec:4_1}
We grouped the magnetars' properties into timing and emission properties for a cross-correlation study.
The six timing properties, $P$ and $\dot P$, $B_{\rm S}$, $\dot E_{\rm SD}$, ${\tau_{c}}$, and ${\dot{\nu}}$
are all derived from two measured parameters: $P$ and $\dot P$.
The number of emission properties differs for each spectral model (Table~\ref{ta:ta2}), but 
we used the following 9 properties: $kT_1$, $R_{\rm 1,BB}$,
$L_{\rm BB}$ ($=L_{\rm 1,BB}+L_{\rm 2,BB}$), $\Gamma$, PL flux ($F_{\rm PL}$),
PL luminosity ${L_{\rm PL}}$, total X-ray luminosity $L_{\rm X}$ (=$L_{\rm BB}+L_{\rm PL}$),
a distance independent luminosity ratio $\zeta=L_{\rm PL}/L_{\rm BB}$, and $\eta$.
Note again that $F_{\rm PL}$, ${L_{\rm PL}}$, $\eta$, and $\xi$ were measured in the 0.5--10\,keV band.

\begin{figure*}
\centering
\includegraphics[width=170mm]{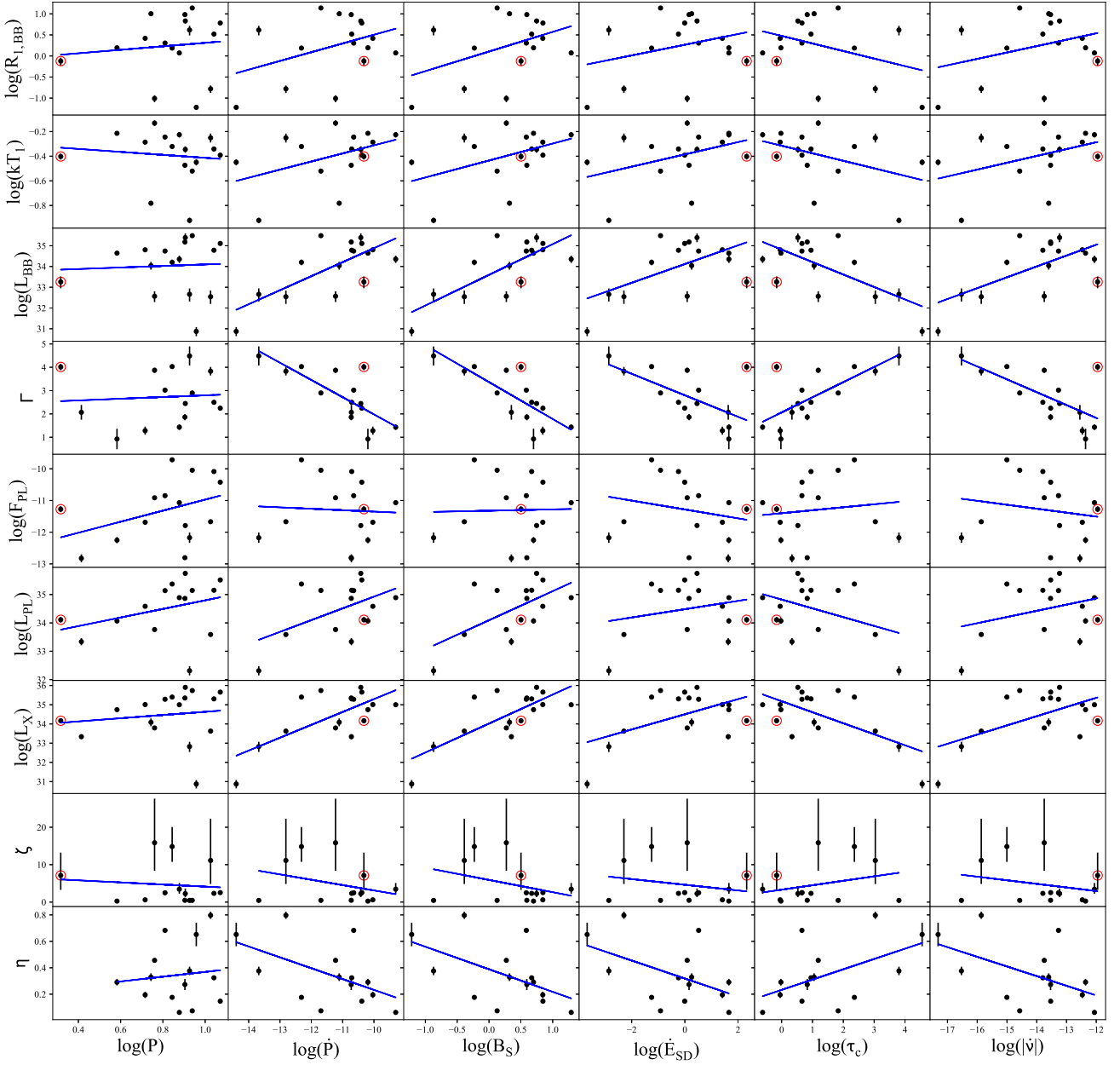}\\
\caption{Scatter plots of the emission and temporal properties. Least-square fits of the correlation trends with linear functions are displayed in blue for reference. Parameters for the magnetar 1E~1547.0$-$5408 are denoted by red circles (see Section~5.1).
\label{fig:fig3}}
\end{figure*}

\begin{figure*}
\centering
\vspace{0mm}
\begin{tabular}{lc}
\includegraphics[width=140mm]{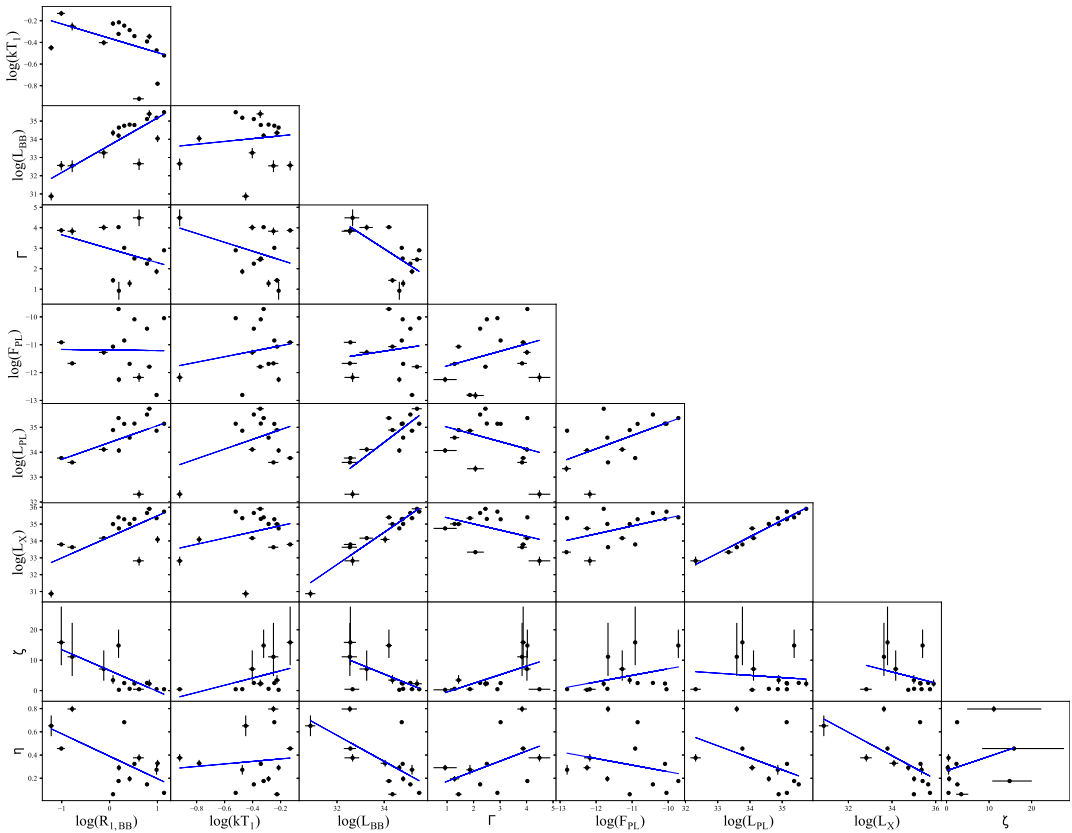}\\
\end{tabular}
\caption{Scatter plots of the emission properties. The best-fit linear functions are displayed in blue for reference.
\label{fig:fig4}}
\end{figure*}

We constructed  54 temporal-emission property pairs and
36 emission-emission property pairs.
Since power-law relations between some properties have been predicted by magnetar models,
we employed a log scale for the
properties except for $\Gamma$, $\eta$,  and $\zeta$  \citep[see also][]{Kaspi2010,Enoto2010,An2012,Mong2018}.
Scatter plots of the property pairs are displayed in Figures~\ref{fig:fig3} and \ref{fig:fig4}.
For each pair of the properties, we computed the Pearson's correlation coefficient and
used the Fisher transformation \citep{Fisher1915} to estimate the correlation
significance ($\kappa_1$). Property pairs with $|\kappa_1|\ge 3.0$ or $|\kappa_2|\ge 3.0$ (Section~\ref{sec:4_2}), and their correlation significances are presented in Table~\ref{ta:ta2}.

%%% FIGURE %%%%%%%%%%%%%%%%%%%%%%%%%%%%%%%%%%%%%%%%%%%%%%%%%%%%%%%%%%%%%%%%%%%%
%%%%%%%%%%%%%%%%%%%%%%%%%%%%%%%%%%%%%%%%%%%%%%%%%%%%%%%%%%%%%%%%%%%%%%%%%%%%%%%

%%% FIGURE %%%%%%%%%%%%%%%%%%%%%%%%%%%%%%%%%%%%%%%%%%%%%%%%%%%%%%%%%%%%%%%%%%%%
%%%%%%%%%%%%%%%%%%%%%%%%%%%%%%%%%%%%%%%%%%%%%%%%%%%%%%%%%%%%%%%%%%%%%%%%%%%%%%%

\setlength\tabcolsep{1pt}
%\begin{landscape}
\begin{tiny}
%%% Table %%%%%%%%%%%%%%%%%%%%%%%%%%%%%%%%%%%%%%%%%%%%%%%%%%%%%%%%%%%%%%%%%%%%%%%
\begin{table}
\centering
\caption{Significantly correlated property pairs}
\label{ta:ta2}
\scriptsize{
\begin{tabular}{cc|ccc|ccc}
%\toprule
\hline
\multicolumn{2}{c|}{Properties}& $\kappa_1^{\rm a}$ & $\langle \kappa_1\rangle^{\rm a}$ & $\sigma_{\kappa 1}^{\rm b}$ & $\kappa_2^{\rm a}$ & $\langle \kappa_2\rangle^{\rm a}$ & $\sigma_{\kappa 2}^{\rm b}$  \\
	& 	& ($\sigma$)	& ($\sigma$) & ($\sigma$)	& ($\sigma$) & ($\sigma$) &  ($\sigma$)      \\ \hline
$\dot P$	& $L_{\rm BB}$	&3.3		& 3.2	& 0.3	& 3.3	& 3.3	& 0.3   \\
$\dot P$	& $\Gamma$	&$-$3.4	 & $-$3.4	& 0.4	&$-$2.5	& $-$2.6	& 0.3  \\ 
$\dot P$	& $L_{\rm X}$	&3.3		& 3.0	& 0.3 	&3.2	& 2.9	& 0.3  \\ 
$B_{\rm S}$	& $L_{\rm BB}$	&3.6		& 3.5	& 0.3 	&3.6	& 3.7	& 0.4  \\ 
$B_{\rm S}$	& $\Gamma$	&$-$3.6     & $-$3.5	& 0.5 	&$-$2.7	& $-$2.7	& 0.4   \\ 
$B_{\rm S}$	& $L_{\rm X}$	&3.7		& 3.4	& 0.3 	&3.6	& 3.3 	& 0.3   \\ 
$\tau_{\rm c}$	& $L_{\rm BB}$	&$-$3.0		& -2.9	& 0.3 	&$-$3.0	& $-$3.0 & 0.3   \\ 
$\tau_{\rm c}$	& $\Gamma$	&3.1		& 3.1	& 0.4 	&2.2	& 2.3 & 0.2  \\ \hline
$R_{\rm 1,BB}$	& $L_{\rm BB}$	&4.2		& 4.2	& 0.3 	&4.1 	& 4.0    & 0.3  \\
$R_{\rm 1,BB}$	& $L_{\rm X}$	&3.0	& 2.9 	& 0.3 	&2.9 	& 2.8   & 0.3   \\
$R_{\rm 1,BB}$	& $\zeta$	&$-$3.4		& $-$2.8	& 0.5	&$-$1.8	& $-$1.4 & 0.5   \\
%\JSG{$R_{\rm 1,BB}$}	& \JSG{$\eta$}	&$-$2.5 	& $-$3.0	& 0.4 	&$-$2.5	&	$-$3.2 & 0.6  \\
$L_{\rm BB}$	& $L_{\rm PL}$	&3.9		& 4.0	& 0.4	&2.5	& 2.7   & 0.5  \\
$L_{\rm BB}$	& $L_{\rm X}$	&6.5		& 6.3	& 0.4	&6.5	& 6.3	& 0.4   \\
%\JSG{$L_{\rm BB}$}	& \JSG{$\eta$}	&$-$2.6	 & $-$3.3	& 0.4 	&$-$2.6	&$-$3.5	& 0.4 \\
$F_{\rm PL}$	& $L_{\rm PL}$	&2.1		& 2.2	& 0.3   	&3.9	&  3.6 & 0.9   \\
$F_{\rm PL}$	& $L_{\rm X}$	&1.8		& 1.9	& 0.3 	&3.1	& 3.0  & 0.7  \\
$L_{\rm PL}$	& $L_{\rm X}$	&7.4		& 7.7	& 0.6    	&5.5	& 6.0   & 0.6  \\ \hline  
%\bottomrule
\end{tabular}}
\tabnote{
$^{\rm a}$ Correlation significance in units of $\sigma$. Negative values mean anti-correlation.\\
$^{\rm b}$ Variation of the correlation significance due to uncertainties on the properties. See text for more details}
\end{table}
%%%%%%%%%%%%%%%%%%%%%%%%%%%%%%%%%%%%%%%%%%%%%%%%%%%%%%%%%%%%%%%%%%%%%%%%%%%%%%%

\end{tiny}
%\end{landscape}

Note that it is difficult to accurately measure $P$ and $\dot P$ of magnetars because of large
timing noise and glitches. Moreover,
distances ($d$'s) to magnetars are poorly constrained.
Thus there are various suggestions for those
values.\footnote{http://www.physics.mcgill.ca/$\sim$pulsar/magnetar/TabO2.html}
In general, the suggested values of $P$ for a magnetar differ only little ($\Delta P/P\le 10^{-2}$),
which is not a concern for the correlation study.
However, those of $\dot P$ (and $d$) are substantially different.
To take this into account, we performed simulations.
For each target, we randomly picked a $\dot P$ and a $d$ (uniform priors) value among the suggested ones,
and further varied the spectral parameters.
Note that the spectral parameters are known to covary, and the covariance was taken into account in the simulations as follows. We varied $N_{\rm H}$ using the value and uncertainty presented in Table~\ref{ta:ta1}. We held $N_{\rm H}$ fixed at the varied value and refit the spectral data to derive the best-fit spectral parameters appropriate for the varied $N_{\rm H}$. We then used the covariance matrix provided by XSPEC to generate a simulated set of the spectral parameters for each target, and computed the correlation coefficients.
We repeated this procedure 10,000 times and measured the mean ($\langle \kappa_1 \rangle$)
and standard deviation ($\sigma_{\kappa 1}$); these are presented in Table~\ref{ta:ta3} for reference.

The timing properties $\dot P$ and $B_{\rm S}$ are well correlated with
the emission properties $L_{\rm BB}$, $\Gamma$ and $L_{\rm X}$.
$\tau_{\rm c}$ also shows correlations with $L_{\rm BB}$ and $\Gamma$.
Some correlations between the emission properties were anticipated
from the constructions of the quantities (e.g., $L_{\rm X}=L_{\rm BB}+L_{\rm PL}$), but the $R_{\rm 1, BB}$-$L_{\rm BB}$, and $L_{\rm BB}$-$L_{\rm PL}$
correlations are intriguing. The $R_{\rm 1, BB}$-$\zeta$ correlation is also intriguing, but it is induced by a few high-$\zeta$ points with very large uncertainties (Fig.~\ref{fig:fig4}) and thus varies significantly (e.g., $\sigma_{\kappa 1}$ and $\kappa_2$ in Table~\ref{ta:ta2}). Hence this correlation is rather uncertain.

\subsection{Correlation study with alternative spectral models}
\label{sec:4_2}
In Section~\ref{sec:4_1}, we used the spectral parameters of
the baseline models reported in Table~\ref{ta:ta1} (mostly BB+PL).
However, 2BB models have also been suggested and favored for some magnetars
on statistical or physical grounds (see Section~\ref{sec:3_3}).
These targets are listed in Table~\ref{ta:ta3}.
We replaced the BB+PL spectral parameters with the 2BB ones for these targets and computed the correlation significances ($\kappa_2$; Table~\ref{ta:ta2}). We also
performed 10,000 simulations as was done is Section~\ref{sec:4_1},
and measured average correlation significances ($\langle \kappa_2 \rangle$)
and standard deviations ($\sigma_{\kappa 2}$).
The results are reported in Table~\ref{ta:ta2}.

The significantly-correlated properties found using the baseline models mostly
remain unchanged, but some of the PL-related correlations ($\dot P$-$\Gamma$, $B_{\rm S}$-$\Gamma$, and $L_{\rm BB}$-$L_{\rm PL}$)
became less significant when the 2BB parameters were used
perhaps because of the reduction in the number of PL samples.
Notice that those PL-related correlations
are still modest (e.g., $|\kappa_2|\geq$2.5 except for the $\tau_{\rm c}$-$\Gamma$ correlation). Correlations with the higher-temperature BB properties
($R_{\rm 2,BB}$ and $L_{\rm 2,BB}$) could not be measured due to the lack of 2BB samples.

%%% Table %%%%%%%%%%%%%%%%%%%%%%%%%%%%%%%%%%%%%%%%%%%%%%%%%%%%%%%%%%%%%%%%%%%%%%%
\begin{table}
\centering
\caption{2BB parameters for some magnetars}
\label{ta:ta3}
\scriptsize{
\begin{tabular}{lllcccc}
%%%\toprule
\hline
Name & Model  & $kT_1$ & ${R_{1,\rm BB}}$ & $kT_2$ & ${R_{2,\rm BB}}$  \\
 &        & (keV)     & (km)          & (keV)             & (km)   \\ \hline
J0100  & 2BB    & 0.29(2)  & 12(1)  & 0.57(7)  &  2(1) \\ 
SGR0501 & 2BB & 0.27(1)  & 1.5(1) & 0.72(2) & 0.15(1)  \\ 
SGR0526 & 2BB & 0.40(2) & 11(1) & 1.0(1) & 1.2(4)  \\ 
J1822  & 2BB & 0.11(1) & 6(1) & 0.27(2) & 0.3(1) \\\hline
\end{tabular}
}
%\tabnote{
\footnotesize{
}
\end{table}

\section{Discussion}
\label{sec:5}
We presented a refined list of quiescent magnetars and
investigated correlations between their spectral and temporal properties, including
$\eta$  which has not been explored previously. We further considered uncertainties in the spectral
and temporal parameters as well as the distances to the targets using simulations. Hence, our correlation study is more thorough than the previous ones. 
Here we discuss some intriguing correlations (e.g., $|\kappa_1|\gtrsim 3$; Section~\ref{sec:4_1}) obtained using the baseline models (Table~\ref{ta:ta1}). Using 2BB models for some of the targets (Section~\ref{sec:4_2}) alters significances of PL-related correlations as noted above.

\subsection{Significant correlations between temporal and emission properties}
\label{sec:5_1}
As noted in Section~\ref{sec:4_1}, the timing properties are derived from $P$ and $\dot P$
measurements, and then the correlations of the emission properties with $\dot P$
(Table~\ref{ta:ta2}) might have induced those with the other timing properties
that are combinations of $P$ and $\dot P$ (e.g., $B_{\rm S}$ and $\tau_{\rm c}$).
The timing properties are thought to represent physical quantities approximately, and
the results (Table~\ref{ta:ta3}) show that
there are better (or worse) combinations of $P$ and $\dot P$ that make
the correlations more (less) significant.
These can provide insights into magnetars' evolution and emission mechanisms \citep[e.g.,][]{Marsden2001,Kaspi2010,An2013a,Mong2018}.

The results presented in Table~\ref{ta:ta2} confirm the previously suggested
$B_{\rm S}$-$\Gamma$ \citep[][]{Kaspi2010} and
$B_{\rm S}$-$L_{\rm X}$ \citep[][]{An2012,Mong2018} correlations.
The $B_{\rm S}$-$\Gamma$ (anti-)correlation has been explained as due to an increased optical depth to the
RCS in high-$B$ sources \citep[][]{Thompson2002};
photons make multiple scattering and the magnetospheric particles have larger velocity spread,
rendering the soft nonthermal spectrum harder \citep[e.g.,][]{Fernandez2007,Kaspi2010}.
For the $B_{\rm S}$-$L_{\rm X}$ correlation, we further separately investigated the thermal ($L_{\rm BB}$) and
nonthermal ($L_{\rm PL}$) luminosities,
and found that $B_{\rm S}$ is significantly correlated with $L_{\rm BB}$ and not with $L_{\rm PL}$ ($\kappa_1=2.4$). This indicates that the $B_{\rm S}$-$L_{\rm X}$ correlation is mainly driven by
the $B_{\rm S}$-$L_{\rm BB}$ one.
We speculate that this may be because $L_{\rm BB}$ is directly affected by the $B$ decay, whereas $L_{\rm PL}$ is influenced by both $L_{\rm BB}$ (seeds for RCS) and
magnetospheric current (scatterer for RCS). Hence the $B_{\rm S}$-$L_{\rm PL}$ correlation might have been blurred.
The $B_{\rm S}$-$L_{\rm X}$ correlation supports the idea that magnetars' emission power is supplied primarily
by the internal $B$ decay for which magnetar models predict
$L_{\rm X}\propto B^{4.4}$ \citep[original magnetar model;][]{Thompson1996}
or $L_{\rm X}\propto B$ \citep[twisted-$B$ model;][]{Thompson2002}.
In our fit of the $B_{\rm S}$-$L_{\rm X}$ trend (Fig.~\ref{fig:fig3}),
we found $L_{\rm X}\propto B_{\rm S}^{1.5}$ (and $L_{\rm BB}\propto B_{\rm S}^{1.5}$).
While the spin-inferred $B_{\rm S}$'s may not accurately represent true $B$'s of magnetars,
if assuming so, the result seems to agree reasonably with the twisted-$B$ model.

We discovered intriguing correlations of $\tau_{\rm c}$ with $L_{\rm BB}$ and $\Gamma$.
These correlations are expected as magnetars lose their internal $B$
energy via long-term cooling; $L_{\rm BB}$ drops and the nonthermal (RCS) emission softens with time. Theoretically,
\citet[][]{Thompson1996} predicted $L_{\rm BB}\propto t^{-0.3}$--$t^{-0.4}$ and a recent
magneto-thermal evolution model predicted diverse trends
depending on the initial configuration of $B$ \citep[e.g., Fig.~7 of][]{Vigano2013}.
To compare with the model predictions,
we fit the $\tau_{\rm c}$-$L_{\rm BB}$ trend with a power-law function
and found $L_{\rm BB}\propto \tau_{\rm c}^{-0.6}$.
It appears to agree reasonably well with the theoretical predictions of \citet[][]{Thompson1996} and \citet{Vigano2013}.

\citet{Marsden2001} suggested a possible correlation between the spin-down torque ($\dot \nu$)
and $\Gamma$ using a small sample (7 magnetars).
\citet{Kaspi2010} investigated this correlation using a larger sample (11 magnetars)
and noted that the $\dot \nu$-$\Gamma$ correlation is significant only at 1.8$\sigma$ which increases to 2.7$\sigma$ when ignoring 1E~1547.0$-$5408 for which
the $\Gamma$ measurement was very uncertain \citep[i.e., $\Gamma=3.7^{+0.8}_{-2.0}$;][]{Gelfand2007}.
Later \citet[][]{Bernardini2011} refined the $\Gamma$ measurement ($\Gamma=4.0\pm0.2$)
using multi-epoch data. Hence we did not exclude 1E~1547.0$-$5408 from
our correlation study (15 magnetars) and
found a modest correlation between $\dot \nu$ and $\Gamma$ at 2.8$\sigma$.
This correlation is certainly intriguing but is not yet definitive.
Related to the $\dot \nu$-$\Gamma$ correlation, we found that
the spin-down `rate' $\dot P$ ($-\frac{\dot \nu}{\nu^2}$)
is better correlated with
$\Gamma$. This correlation is in accordance with the twisted-$B$ model of magnetars
\citep[][]{Thompson2002} which predicted that the twist increases
the current flowing across the light cylinder, thereby resulting in
an increase of the spin-down rate and the optical depth; the latter
makes the spectrum harder (i.e., smaller $\Gamma$).

1E~1547.0$-$5408 seems to be an outlier for the $\Gamma$-related correlations (Fig.~\ref{fig:fig3})
as was noted by \citet{Kaspi2010} for the $\dot \nu$-$\Gamma$ correlation.
Ignoring it from the sample makes all the $\Gamma$-related
correlations stronger. 
We note that the property pairs of the source, except for the $\Gamma$-related ones, lie close to the correlation trends (Fig.~\ref{fig:fig3}), meaning that the source has properties of typical magnetars but its nonthermal emission is very soft. It is intriguing to note that the other target XTE~J1810$-$197 also shows extremely soft PL ($\Gamma=7$) emission for its rotational properties,
if we consider the BB+PL model for the source (Section~\ref{sec:3_3}). Magnetospheric X-ray emission of these targets seems to be highly suppressed despite their strong $>10^{14}$\,G field. 
This may be related to the fact that the two magnetars 1E~1547.0$-$5408 and XTE~J1810$-$197 are the few `radio magnetars' in which pulsed radio signals have been detected \citep[only two in our target list; see also][]{Chu2021}. This is only speculative, and further studies are needed to address this issue.

\subsection{Correlation between $B_{\rm S}$ and $kT$}
\label{sec:5_2}
As we showed above (Section~\ref{sec:5_1}), $B_{\rm S}$ has significant
influence on BB emissions of magnetars. Then, $kT_{\rm BB}$ may also
be correlated with $B_{\rm S}$. However, we did not find any significant
correlation between them ($\kappa_1=1.7$).
Note that \citet{Pons2007} discovered a $B_{\rm S}$-$kT_{\rm BB}$ correlation
in a sample of `isolated neutron stars' with
$B_{\rm S}$ in the range of $10^{12}$--$10^{15}$\,G (i.e., including some thermally emitting X-ray pulsars and magnetars).
They further measured the correlation trend to be $kT_{\rm BB}\propto B_{\rm S}^{0.5}$
and suggested that the correlation implies that the
internal heat is generated by magnetic field decay: $B_{\rm S}^2\propto T_{\rm BB}^4$. However, it appears that the correlation in the magnetar group alone seemed insignificant in that work \citep[Fig.~1 of][]{Pons2007}, and \citet{Zhu2011} noted that the $B_{\rm S}$-$kT_{\rm BB}$ correlation is insignificant in a larger sample of neutron stars (normal pulsars, high-$B$ pulsars, and X-ray-isolated neutron stars) than was used by \citet{Pons2007}.
These mean that the $B_{\rm S}$-$kT_{\rm BB}$ correlation suggested by \citet{Pons2007} might be caused by clustering of the pulsars (low $B_{\rm S}$ and $kT_{\rm BB}$) and the magnetars (high $B_{\rm S}$ and $kT_{\rm BB}$): i.e., a correlation between the pulsar and the magnetar populations not within the ``isolated neutron star'' population.

In the magnetar population, \citet{Rea2008}, using a physically-motivated RCS model, found no significant correlation between the surface temperature ($kT$)
and $B_{\rm S}$. On the other hand, \citet{Mong2018} used a 2BB or a 2BB+PL model for a sample of magnetars, and suggested that the model-inferred temperature for the cooler BB emission with a radius greater than 3\,km is correlated with $B_{\rm S}$ ($kT_{\rm BB}\propto B_{\rm S}^{0.4}$).
So the current situation for the $B_{\rm S}$-$kT_{\rm BB}$ correlation in the magnetar population is unclear as the correlation significance alters substantially
depending on the spectral model and targets used for the studies. Further theoretical and observational studies are warranted. 

\subsection{Correlations between emission properties}
\label{sec:5_3}
We found significant correlations in a number of emission property pairs
(Table~\ref{ta:ta2} and Fig.~\ref{fig:fig4}), which can provide hints to emission mechanisms of magnetars as compared to RPPs. As we noted above, both magnetars and RPPs emit nonthermal X-ray radiation in the magnetosphere, but suggested emission mechanisms are very different: RCS off of the thermal seed photons (i.e., $L_{\rm BB}$) for magnetars {\it vs} synchrotron radiation for RPPs. While the RCS scenario has been favored over the synchrotron scenario for magnetars based on theoretical arguments and observed spectral features (e.g., spectral turn-over at $\gtrsim$10\,keV in some magnetars), further observational supports would help to discern the scenarios more clearly. A difference between the RCS and synchrotron scenarios is that the nonthermal emission is strongly affected by the thermal one as the latter provides seeds for the former in the RCS scenario, whereas the thermal and nonthermal emissions are not strongly related to each other in the synchrotron scenario. Hence, the correlation we found between $L_{\rm BB}$ and $L_{\rm PL}$ supports the RCS scenario for magnetar's nonthermal emission. 

The $R_{\rm 1, BB}$-$L_{\rm 1,BB}$ (and $R_{\rm 1,BB}$-$L_{\rm X}$) correlation is observationally obvious as $L_{\rm BB}\propto R_{\rm BB}^{2} kT_{\rm BB}^4$. In this case, however, the lack of $L_{\rm BB}$-$kT_{\rm BB}$ correlation ($\kappa_1=0.4$) is puzzling. This is probably because the simple BB+PL or 2BB model only approximately represents magnetars' emission. For example, some low-energy contamination from multiple cold spots and/or magnetosphere might be ascribed to the BB model, which would increase the fit-inferred $R_{\rm 1,BB}$ and $L_{\rm BB}$ but lower $kT_{\rm BB}$, thereby enhancing the $R_{\rm 1,BB}$-$L_{\rm BB}$ correlation and blurring the $kT_{\rm BB}$-$L_{\rm BB}$ one.

\subsection{Comparisons with other neutron-star populations}
\label{sec:5_4}
Correlations of the emission and temporal properties in the populations of RPPs with nonthermal X-ray emission \citep[][]{Li2008} and thermally emitting pulsars \citep[][]{Zhu2011} have also been studied.
\citet{Li2008} carried out a correlation study with temporal and `nonthermal' emission properties of 27 RPPs. Although only some of the property pairs are correlated with high significance (e.g., chance probabilities of $p<10^{-3}$), it appears that the RPPs' nonthermal X-ray luminosity $L_{\rm X,psr}$ is well correlated with their temporal properties $P$, $\dot P$, $\tau_{\rm c}$, and $\dot E_{\rm SD}$, and X-ray photon index $\Gamma$ is reasonably well correlated with $P$ and $\dot E_{\rm SD}$. Neither $L_{\rm X,psr}$ nor $\Gamma$ was found to be correlated with $B_{\rm S}$; we confirmed this by reanalyzing the data presented in \citet{Li2008}. Recalling that magnetars' $\Gamma$ and $L_{\rm X}$ are correlated with $B_{\rm S}$ but not with $\dot E_{\rm SD}$ (Table~\ref{ta:ta2}), the two populations, magnetars and RPPs, seem to be very different. This supports the idea that the primary energy sources of magnetars and RPPs are different; $B$ for the former and the rotational energy for the latter.

On the other hand, \citet{Zhu2011} investigated correlations of `thermal' emission properties with temporal ones using a sample of thermally emitting pulsars \citep[high-$B$ pulsars, normal pulsars and X-ray-isolated neutron stars; see][for more detail]{Zhu2011}. Because the authors considered only $B_{\rm S}$-$kT$ correlation, we reanalyzed the data ($kT$, $R_{\rm BB}$, and $L_{\rm BB}$) presented in \citet{Zhu2011} after supplementing them with $P$ and $\dot P$ taken from the ATNF catalog\footnote{https://www.atnf.csiro.au/research/pulsar/psrcat/}, and found out that $L_{\rm BB}$ is correlated with $\tau_{\rm c}$ at the 3$\sigma$ level. Again, neither $L_{\rm BB}$ nor $kT$ is significantly correlated with $\dot E_{\rm SD}$ or $B_{\rm S}$ having significances $<2\sigma$. This may indicate that cooling of these thermally emitting pulsars occurs primarily by release of the residual heat in the core, not by the spin down or a $B$ decay. The cooling trend of the sources is measured to be $L_{\rm BB}\propto \tau_{\rm c}^{-0.47}$. These pulsars seem to cool relatively slowly compared to magnetars ($L_{\rm BB}\propto \tau_{\rm c}^{-0.6}$), meaning that magnetars' energy loss (by $B$ and residual heat) is larger than the thermally emitting pulsars'.

In summary, these comparisons suggest that thermal and nonthermal emissions of the pulsars (i.e., RPPs and thermally emitting pulsars) arise from residual heat and spin-down energy, respectively, whereas magnetars' emission is strongly affected by the decay of $B$.
While there are more to be studied by analyzing emission and temporal properties of the neutron-star populations (i.e., magnetars, RPPs, and the thermally emitting pulsars) simultaneously, we defer such a research to future work, since we need to scrutinize the measurements made for the pulsars \citep[e.g.,][]{Li2008, Zhu2011} presumably by reanalyzing the data as we did here for magnetars.

\section{Summary}
\label{sec:6}

\begin{itemize}
\item We found that the emission properties, the thermal luminosity
$L_{\rm BB}$ and the X-ray photon index $\Gamma$,
are correlated with the spin-down rate ($\dot P$), the surface dipole magnetic
field strength ($B_{\rm S}$) and characteristic age $\tau_{\rm c}$.

\item We found $L_{\rm BB}\propto B_{\rm S}^{1.5}$ and
$L_{\rm BB}\propto \tau_{\rm c}^{-0.6}$ trends
which are similar to predictions of magnetar models.

\item We found that $L_{\rm BB}$ is correlated with $L_{\rm PL}$. This correlation supports the RCS scenario for magnetars' nonthermal emission.

\item The correlations in the magnetar population are different from those seen in other neutron-star populations (e.g., RPPs and thermally emitting pulsars), indicating that the energy sources for emissions of magnetars and the other neutron stars are different.
\end{itemize}

While the results obtained from our study
suggest intriguing correlations that can help to delineate emission mechanisms in magnetars,
there are things to be improved.
While the spectral models we used in this work seem to represent the thermal and nonthermal properties of magnetars well, some residuals are noticeable (Fig.~\ref{fig:fig1}), possibly suggesting that the actual emission of magnetars may be different from these simple models. A well-justified physical model of magnetars' emissions is lacking, but a correlation study with such a model in the future may reveal different correlations among the physical properties of the stars.
The current identifications of the `quiescent' state of the targets may not be
very accurate. Besides, $\dot P$ of some magnetars
was measured during an outburst period, and in this case
the measurement might be biased to a larger value by a putative
glitch and its recovery \citep[e.g.,][]{Woods2007}.
Better identification of the quiescent states and more accurate measurements of $\dot P$
can be achieved with a deeper and high-cadence monitoring campaign for the magnetars over a long period.
The current/future X-ray missions eROSITA \citep[][]{Predehl2021}, Lynx \cite[][]{Gaskin2019}, AXIS \cite[][]{Mushotzky2019}
and Athena \citep[][]{Barcons2017} will certainly be very helpful.

Some magnetars exhibit a distinct hard X-ray (e.g., $\ge$10\,keV) spectrum which we did not consider because they have been accurately measured only for
a small number of magnetars \citep[e.g.,][]{Kuiper2006}. The hard-band properties may be correlated with the temporal and soft-band properties \citep[e.g.,][]{Thompson2005,Beloborodov2013,Wadiasingh2018}, and 
can provide further insights into magnetar physics \citep[e.g.,][]{Kaspi2010,Enoto2010,Yang2016}.
Deeper NuSTAR \citep[][]{Harrison2013} and future HEX-P \citep[][]{Madsen2018} observations of quiescent magnetars are warranted.

%%% ACKNOWLEDGMENTS (IF ANY) %%%%%%%%%%%%%%%%%%%%%%%%%%%%%%%%%%%%%%%%
\acknowledgments
This research was supported by the National Research Foundation of Korea (NRF) grant
funded by the Korean Government (MSIT) (NRF-2022R1F1A1063468 and NRF-2023R1A2C1002718).

%%% APPENDICES (IF ANY) %%%%%%%%%%%%%%%%%%%%%%%%%%%%%%%%%%%%%%%%%%%%%
%\newpage
%\appendix

%%% PUT YOUR REFERENCES HERE %%%%%%%%%%%%%%%%%%%%%%%%%%%%%%%%%%%%%%%%
%\bibliographystyle{apj}
%\bibliography{magnetars}
%%% END LIST OF REFERENCES %%%%%%%%%%%%%%%%%%%%%%%%%%%%%%%%%%%%%%%%%%
%\end{thebibliography}

%%% APPENDICES (IF ANY) %%%%%%%%%%%%%%%%%%%%%%%%%%%%%%%%%%%%%%%%%%%%%
%\newpage
%\appendix

%%% PUT YOUR REFERENCES HERE %%%%%%%%%%%%%%%%%%%%%%%%%%%%%%%%%%%%%%%%

\end{document}